\documentclass[iop]{emulateapj}

\usepackage{graphicx}
\usepackage[caption=false]{subfig}

\slugcomment{Accepted for publication in the AJ.}

\shorttitle{Giant's MMRs replenishment by DPs}
\shortauthors{Mu\~noz-Guti\'errez et al.}

\begin{document}


\title{The Contribution of Dwarf Planets to the Origin of Low 
Inclination Comets by the Replenishment of Mean Motion Resonances in Debris Disks.}


\author{M. A. Mu\~noz-Guti\'errez}
\affil{Institute of Astronomy and Astrophysics, Academia Sinica, National Taiwan University, Taipei, Taiwan}
\email{mmunoz@asiaa.sinica.edu.tw}
\author{A. Peimbert\, and B. Pichardo.}
\affil{Instituto de Astronom\'ia, Universidad Nacional Aut\'onoma de 
M\'exico, Apdo. postal 70-264 Ciudad Universitaria, M\'exico}



\begin{abstract}

In this work we explore a new dynamical path for the delivery of low-inclination comets. In a configuration formed by an interior giant planet and an exterior massive debris disk, where the mass is accounted for by the 50 largest objects in the disk, the strongest mean motion resonances of the giant, located along the belt, are replenished with new material (test particles) due to the influence of the 50 massive objects. Once in resonance, slow chaotic diffusion stirs the orbital elements of the cometary nuclei enough to encounter the giant and to be scattered by it. When the disk is massive enough, both resonant and non-resonant particles are stirred quickly to encounter the giant and form an scattered disk component, greatly increasing the rate for the delivery of cometary material to the inner part of the system. This mechanism is applicable both to the solar system and extrasolar systems in general. Preliminary results, using a disk as massive as the classical Kuiper belt, indicate that the mechanism here proposed can account for about a tenth of the required injection rate to maintain the population of ecliptic comets in steady state. In a more massive belt of 0.25 M$_\oplus$, an estimated rate of around 0.6 new comets per year is found. Such a high rate would pose a serious risk for the habitability of rocky interior planets, yet would resemble the late heavy bombardment that was present in the early solar system.

\end{abstract}


\section{Introduction}

Beyond the orbit of the outermost planet known of the solar system, a
large reservoir of cometary nuclei, going from the trans-Neptunian
region to the outer edge of the Oort cloud, fills the space up to an incredible distance of approximately 100,000 au. In there, thousands of millions of bodies coexist nearly in steady state. A few of these bodies however are destabilized every now and then producing one of the most staggering astrophysical phenomenologies: comets. A cometary nucleus becomes a comet (i.e. an icy body with a coma and a long tail produced by sublimation of mainly water ice by solar radiation) when its orbit is disrupted, advancing progressively towards the inner solar system, by increasing their eccentricity while reducing their perihelion distance \citep[][and references therein]{Dones15}.

Although the Kuiper belt is far and away the best known cometary reservoir, there are hundreds of planetary systems with conspicuos debris disks or Kuiper belt like structures discovered in the last decade \citep{Wyatt08,Matthews14,Wyatt18b}; some examples of them are the ones circling Fomalhaut, $\beta$Pictoris, Vega, etc. Compared to those debris disks, our Kuiper belt is actually ``anemic'', fortunately for life in the solar system. In spite of its low mass, its existence remains relevant for the inner parts of the solar system and much data has been collected about it. We present here a short overview of the information related to the topic introduced in this work. The solar system data presented below should also be relevant for most extrasolar systems.

In the solar system, once cometary nuclei descend to the inner solar system, they become planet crossers and are exposed to continuous orbital perturbations
that diminish swiftly their survival probabilities; this means that,
given the current population of comets in the inner solar system, the
most probable scenario is that cometary nuclei are continuously
resupplied somehow from the trans-Neptunian region \citep{Fernandez80,Duncan88,Nesvorny17}. 

An important endeavor has been done trying to explain the current
number of comets, particularly the low inclination comets (or ecliptic
comets). In an influential work, \cite{Levison97}, showed that
in order to maintain a constant population of ecliptic comets (defined
by the authors as those with a Tisserand parameter with respect to
Jupiter of $2<T_J<3$), the Kuiper belt objects with orbits that become
Neptune's crossers, could evolve into an orbit with perihelion smaller
than 2.5 au, where they turn into active comets. In that work, the
authors conclude that in order to explain the observed inclination
distribution, the ecliptic comets would need to be active for a long
time (about 12,000 years) after they have reached a 2.5 au perihelion 
distance.  

A first scenario to explain the current population of ecliptic comets
in the solar system is one where the resonant regions of the classical
Kuiper belt are the suppliers \citep[e.g.][]{Morbidelli97,
Tiscareno09}. In that scenario, the gravitational
effect of Neptune is considered as the only one exerted that produces
the escape of bodies from the classical Kuiper belt (CKB), and governs, by
secular chaotic processes in its orbital resonances, the motion of
cometary nuclei in the belt. The low efficiency of such process in
this apparently highly stable environment, made researchers to come up
with new interesting ideas of more efficient processes to explain the
current population of low inclination comets in the inner solar
system. Currently, there seems to be a general consensus that the
classical Kuiper belt is not an important source of the ecliptic
comets because of the extremely stable orbits of the cometary nuclei
after Neptune stopped migrating. However, we will show in this work
that once the effect of dwarf planets is considered, this scenario can
supply a fraction, that under some circumstances might be significant, 
of the observed comets. 

In a second scenario, the in-falling cometary nuclei originate in the
scattered disk instead of in the CKB 
\citep{Duncan97,Volk08}. This scenario takes advantage of the fact that the
scattered disk is far less stable than the classical belt, therefore
it is a more efficient source of ecliptic comets because of their
proximity to Neptune at perihelion \citep[it is also a better source of
Halley type comets][]{Levison06,Volk08,Nesvorny17}. Assuming the scattered disk
is the main source of ecliptic and Halley type comets, estimates of
the required number of cometary nuclei in the scattered disk (with
diameters larger than $\sim2$~km) lie in the range $1-6 \times 10^9$
objects \citep{Levison06,Brasser13,Brasser15,Rickman17}. In this respect, although there are already some constraints, like, for instance, those based on occultation studies \citep{Schlichting09,Bianco10}, we do not have yet a definitive answer. Also, given the wider inclination distribution of cometary nuclei found in the scattered disk \citep[median $\sim20 - 25^\circ$;][]{Gulbis10,Nesvorny16b}, compared to the one of the ecliptic comets \citep[median $\sim 13^\circ$;][]{Nesvorny17}, it is difficult to reconcile this scenario to explain the lowest inclination population of ecliptic comets \citep{Rickman17}. 

\cite{Nesvorny17} present another approach to the solution of
this problem by performing end-to-end simulations from the early stages of the solar system up to 4.5 Gyr \citep[see also][]{Nesvorny12}. Within this model, cometary nuclei reservoirs are formed since the early epochs of the solar system. The number of comets after 4.5 Gyr is proportional to the number of cometary nuclei in the original transplanetary disk, which can be calibrated based on the number of
known Jupiter Trojans \citep{Nesvorny16}, although recent results based on the color discrepancy between the Jupiter and Neptune Trojans sow doubts about a confident and direct relation between these populations and those in the Kuiper belt \citep{Jewitt18}. The authors conclude then that the main source of ecliptic comets is the scattered disk and the main source of Halley type comets is the Oort cloud. 

Of course, although this detailed information and quantification of the comets statistics and direct observation has been up to now only conceivable in the solar system, the astounding discovery of comets in extrasolar planetary systems \citep[e.g.][]{Weaver95,Movshovitz12,Bodman16,Rappaport18,Wyatt18}, known now as exocomets, brings us to an interesting epoch where dynamical studies of debris disks and the origin of exocomets will be closely studied and interpreted in a more statistical sense, as well as individually (including for instance, secular processes that bring comets to the interior regions of planetary systems). 

In this work, we present a study that shows how the secular effect, produced by dwarf planets and minor bodies, on cometary nuclei (specifically on a classical Kuiper belt-like debris disk), becomes influential to the number of comets that descend to an inner planetary system. This mechanism brings cometary nuclei to the resonant regions in a continuous way that ensures an uninterrupted resupply of new comets; in the case of our solar system, this secular effect may be able to explain the origin of up to a fifth (or more) of the yet puzzling ecliptic comets.

This paper is organized as follows: in Section \ref{sec:sims} we describe our debris disk models, the numerical simulations performed, and the result of a massless reference simulation. Section 3 is devoted to present and discuss the results for different disk masses, and the estimations for the in-falling rates of low-inclination comets in extrasolar systems (Sections \ref{Evsep}, \ref{crossers}, \ref{fracEV}, \ref{sec:largeDP}, and \ref{Injection}), including a toy model for the solar system as an example of application in subsection \ref{toymodel}. Finally, in Section \ref{conc} we present our main conclusions.

\section{Methods and Simulations}
\label{sec:sims}

Our simulations were designed with the aim of demonstrating and characterizing the existence of the dynamical processes associated to a giant planet acting on an external debris disk; in this situation dwarf planet-sized objects sometimes play a crucial role on the secular evolution of the disk. While we are trying to study a broad set of physical parameters, in this work we select a toy model that loosely resembles the outer parts of the solar system (Neptune and the cold CKB). At the same time, this model will allow us to expand on it, in order to explore a broader sample of debris disks with giant and dwarf planets. 

We recall that in order to study any specific example, like the solar system itself or any other extrasolar system, a detailed model accounting for the particular conditions of such system (like the mass of the star, mass of the disk, planetary architecture, etc.) would be required; however, such complexity is beyond the scope of the present paper.

In this study we demonstrate the existence of the phenomenology of the replenishment of mean motion resonances (MMRs) by dwarf-planet sized bodies (DPs), we present an initial characterization, as well as the need of a detailed modeling. We do this by obtaining order of magnitude estimates of its potential importance for our toy model of the solar system and for the most massive extrasolar scenarios at hand.

Each one of our simulations consists of a giant planet (equal to
Neptune in all physical and orbital parameters, except for the
inclination that was set to zero), 50 massive DPs, as well as 5000 test particles. All the simulations
include a central star of 1M$_\odot$ and were performed by using the
symplectic integrator contained in the MERCURY package of
\cite{Chambers99}, with an error tolerance of $10^{-10}$
for the Bulirsch-St\"oer integrator with an initial time-step of 180
days, for a total integration time of 1 Gyr.

\subsection{Initial conditions for disk particles.}

Our test particles, initially, are randomly distributed between 38 and 50 au forming a cold debris disk. The distributions of eccentricities and inclinations of the test particles were obtained from two sets of random gaussian distributions of points on a XY plane, from which we extract the eccentricity, $e$, and argument of periastron, $\omega$, in one case, and the inclination, $i$, and longitude of the ascending node, $\Omega$, on the other, as described in \cite{Munoz15}. This method produces random gaussian distributions for $e$ and $i$, with mean and standard deviations $\left<e\right>=0.037$ and $\sigma_e=0.019$ for $e$, and $\left<i\right>=1.52^\circ$ and $\sigma_i=0.80^\circ$ for $i$, respectively. At the same time $\omega$ and $\Omega$ are randomly distributed between 0 and 360$^\circ$. Finally, the mean anomaly, $M$, is assigned randomly between 0 and 360$^\circ$.

\subsection{Initial conditions for the DP distributions.}

\begin{figure}
\plotone{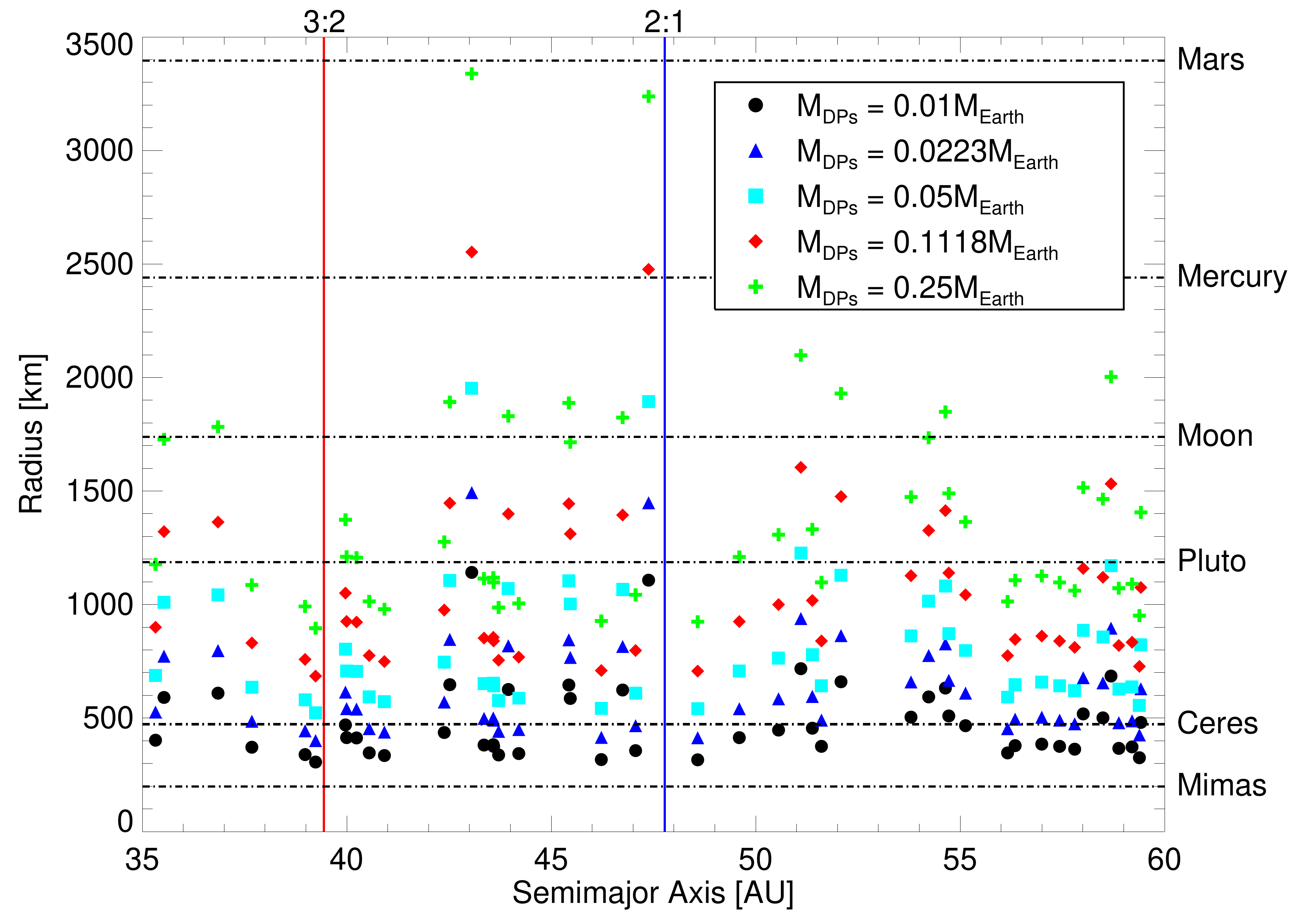}
\caption{Spatial distribution of DPs vs. their size for different disk masses. The radius of each DP in the five different massive disks is plotted against its initial semimajor axis. Note that the spatial location is the same for each single DP, but they are just rescaled in mass to account for progressively more 
massive disks. The names and radius of six representative solar system bodies are indicated by the horizontal black dot-dashed lines and the labels at the right side of the plot. See text for details. Vertical lines mark the location of the 3:2 (red) and 2:1 (blue) MMRs with the giant planet.\label{DPsavsr}}
\end{figure}

\begin{figure}
\plotone{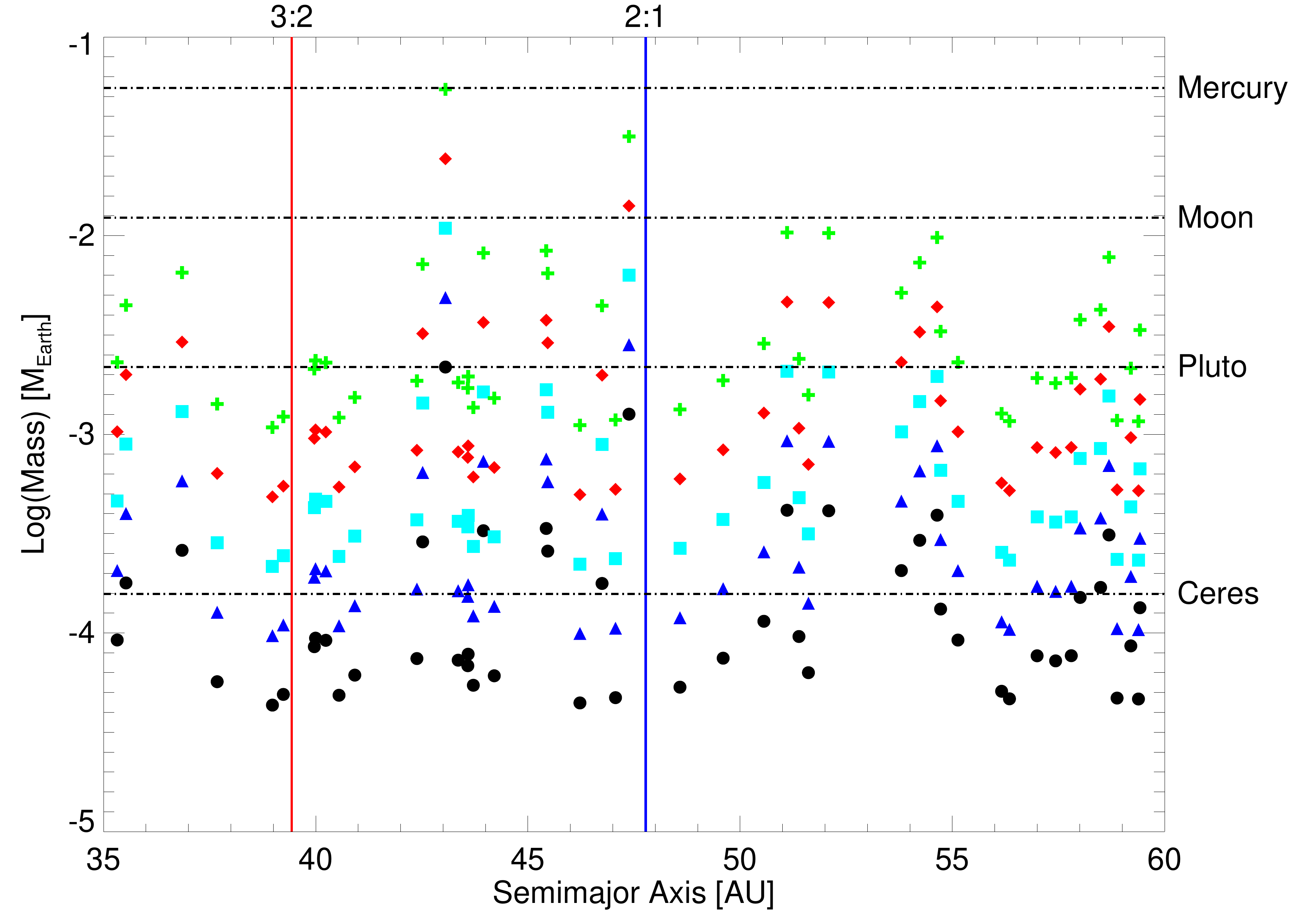}
\caption{Spatial distribution of DPs vs. their mass for different disk total masses. The symbols and reference lines are the same as in Fig. \ref{DPsavsr}. Note how our heaviest DP in the most massive disk is only as massive as Mercury despite being as large as Mars, due to the densities for icy bodies considered in this work.\label{DPsavsm}}
\end{figure}

The DPs are randomly distributed between 35 and 60 au, and they
constitute the massive part of the debris disk, i.e., we assume the
mass of the disk equals the added mass of the 50 DPs. Regarding the
orbital distribution, all eccentricities and inclinations are randomly assigned between 0 and 0.1 and 0$^\circ$ and 5$^\circ$, respectively, while the three angles, $\omega$, $\Omega$, and $M$, are randomly assigned between 0$^\circ$ and 360$^\circ$. The simulations shown in this work included some of those presented previously in \cite{Munoz17}, plus several new cases.

In \cite{Munoz17} we demonstrated that the excitation levels of the
disk particles are weakly dependent on the index of the DP
differential mass distribution (dMD), $\alpha$. For a differential size distribution (dSD) of the form $dN/dD\propto D^{-q}$, a dMD will be given by:
$dN/dM\propto M^{-\alpha}$, where $\alpha$ is related to the index of the dSD, $q$, as $\alpha=(2+q)/3$; since, for a constant density, $D\propto M^{1/3}$. For disks in collisional equilibrium, $\alpha\sim1.8$ ($q\sim3.5$), while for the distribution of the
largest Kuiper belt objects, $\alpha\sim2.2$ ($q\sim4.5$) \citep{Dohnanyi69,Fraser08a,Fraser08b}; based on this, throughout this work we make use of $\alpha=2$ for all our simulations \citep[note that in][we use an $\alpha=1.8$ for the
experiments of Section 3.4]{Munoz17}.
  
We used debris disks of five different masses, as well as a massless case; the masses covered in this work are $(\sqrt 5)^n\times{0.01{\rm M}_{\oplus}}$, where $n$ takes the values
$[0,1,2,3,4]$; for reference, estimates of the mass of the CKB vary between $0.008{\rm M}_{\oplus}<{M_{\it CKB}}<0.06{\rm M}_{\oplus}$ \citep{Trujillo01,Bernstein04,Fuentes08,Vitense10,Fraser14}. These masses correspond approximately to the range covered by our first three models.

Figs. \ref{DPsavsr} and \ref{DPsavsm} show the spatial distribution of all the DPs
against their radius in km and mass in Earth masses, respectively. The radius of the DPs are calculated from
their assigned random masses and densities. As we described in \cite{Munoz17},
the mass of each DP, $m_n$, is assigned using the formula:
\begin{equation}\label{massdist}
m_n=\left(\frac{K_{\alpha}}{(\alpha-1)n'}\right)^{1/(1-\alpha)}.
\end{equation} 
where, $\alpha=2$, while $K_\alpha$ is a scaling constant related to the largest mass permitted in the distribution (e.g., for the 0.01M$_\oplus$ disk, $K_\alpha$ is similar to the mass of Pluto). On the other hand, densities are randomly assigned between 1 and 2.5 g$\,$cm$^{-3}$, to cover the range from objects formed mainly by ice, to the denser dwarf planets in the solar system \citep[Eris and Haumea;][]{Barr16}. In Fig. \ref{DPsavsr} we include six horizontal black dot-dashed lines that indicate the radius of six representative solar system bodies. Mimas is the smallest body in the solar system known to be round due to self-gravity, therefore by analogy, all of our objects, even in the less massive disk, would be spherical, even if they were composed by pure ice; the radius of Ceres imposes a limit for the proper recognition as dwarf planets (according to the current IAU criteria). Therefore, for our less massive disk, 19 of the objects in the distribution would be classified as dwarf planets, 39 in the 0.0223M$_\oplus$ disk, and all the 50 objects for the three most massive disks.

It can be seen from Figs. \ref{DPsavsr} and \ref{DPsavsm} that two objects are
significantly more massive in each distribution than the rest of the
DPs; these two objects will likely dominate the dynamics of the disk
particles. It is worth to note that, for a disk of mass equal to 0.01M$_\oplus$, these two objects are similar to Pluto, both in mass and size, while for the most massive disk (0.25M$_\oplus$), those two objects are almost as large as Mars, but only as massive as Mercury (the density of both, Mercury and Mars, is much larger, approximately 4--5~g$\,$cm$^{-3}$), while the rest of the DPs range in size from a bit smaller than Pluto to a little larger than the Moon, while in mass most of them are only as massive as Pluto.

\subsection{Reference simulation with a ``zero-mass'' disk.}

\begin{figure}
\plotone{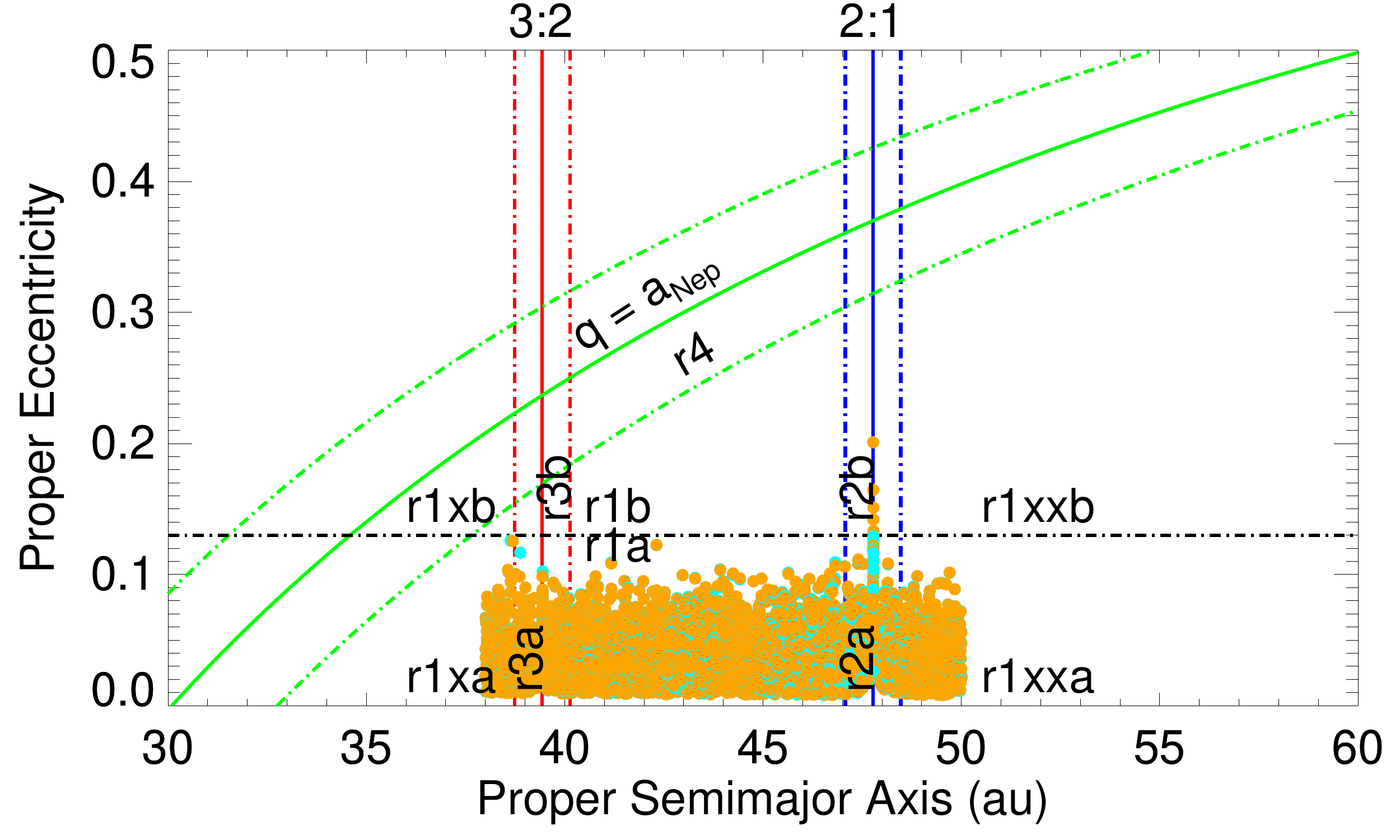}
\caption{Initial and final distributions of proper elements for the reference simulation. We also show the different regions in the plane we have defined to track the evolution of particles. See text for details.\label{refsim}}
\end{figure}

The aim of this work is to explore the evolution of cometary nuclei, initially in the form of a cold debris disk, under the gravitational influence of an interior giant planet, and a distribution of DPs with different masses. In order to better understand the effect of DPs on the cometary nuclei on Gyr time scales, we first establish a point of reference by exploring the ``zero-mass'' disk case; i.e., a debris disk which contains only test particles (and no DPs) under the influence of a Neptune-like giant planet.

To analyze the evolution of test particles, throughout the paper we 
transform the osculating orbital elements ($x(t')$, where $x$ represents $a$, $e$, or $i$) which constitute the usual output of our simulations, to proper orbital elements ($x_p(t)$), namely: the
proper semimajor axis, $a_p$, the proper eccentricity, $e_p$, and the proper inclination, $i_p$. The transformation is performed following the
method described in Section 12.2 of \cite{Morbidelli02}, such that:
\begin{equation} x_p(t)=\frac{1}{N}\sum^{t'=t+5Myr}_{t'=t-5Myr}x(t'),
\end{equation} where $N$ is the number of data points in the 10 Myr running window considered for the averaging procedure; this results in 100 data points of proper elements for each particle for the
full 1 Gyr integration time span. 

The result from the reference simulation is shown in Fig. \ref{refsim}, where cyan dots represent the first set of data points after the averaging procedure (this means they are not exactly equal to the osculating initial conditions of the test particles), while orange dots are the last set of data points, after 1 Gyr of evolution under the influence of the Neptune-like giant. In this phase-space plane of proper orbital elements, $a_p$ vs. $e_p$, we have defined several regions in order to track the evolution of particles:

First, we established a proper eccentricity limit, $e_{lim}$, below which all particles remain at the first data record of the proper elements for the reference simulation. This limit, $e_{lim}=0.13$, is shown as a horizontal black dot-dashed line. Graphically, this limit separates five regions labeled as ``a'' (below the limit) from five regions labeled as ``b'' (above the limit). While physically, this limit separates particles that are mostly unaffected and that we consider to be cold, from those that have been already affected by the several bodies of the system, which we consider to be hot.

The second limit identifies particles that, after some evolutionary
time, are thrown to a region where they are subjected to perturbations
from the giant planet by close encounters. We define this limit by
using the Hill radius of the giant, $R_H=a_{gp}(M_p/3M_\odot)^{1/3}$,
where $a_{gp}$ and $M_p$ are the semimajor axis and mass of the giant
planet. Following \cite{Gladman90,Bonsor12}, we consider that a particle 
interacts with a planet once its periastron is inside the limits given
by $a_{gp}\pm2\sqrt3R_H$. The lines delimiting this region are shown
in the plane $a_p$ vs. $e_p$ by the green dot-dashed lines. We also 
show the line of constant periastron equal to $a_{gp}$
with a solid green line. Region 4 lies above the lower green
dot-dashed line; particles in this region will be labeled as ``crossers''.
The importance of these particles relies on the fact that they are the most likely to quickly evolve into diverse dynamical families: e.g., in the solar system, particles reaching the direct influence
of Neptune could evolve into the scattered disk and all the way 
down into short-period comets \citep{Levison97}. 

These two limits separate the particles that have not been meaningfully affected (regions ``a''), from those that have been significantly heated but have not reached the gravitational influence of the giant planet (regions ``b''), and from those that have reached the gravitational influence of the giant planet (region 4).

We established a third set of limits from the two stronger MMRs present
in the $a_p$ vs. $e_p$ plane considered here, namely the 3:2 and 2:1
MMRs with the giant planet. We define a region limited by $\pm$0.7 au
at the nominal position of the resonances, this is, between 38.7 and 
40.1 au, for the 3:2 MMR, and between 47.0 and 48.4 au, for the 2:1 MMR. 
We adopted this limit following the approach of \cite{Tiscareno09}, who consider a particle to be potentially resonant if its semimajor axis remains within such interval; this width is overgenerous to be sure to include all resonant particles. Those regions are labeled as region 2 and 3, for the 2:1 and 3:2 MMRs, respectively (where 2a and 2b, as well as 3a and 3b regions are labeled depending on the $e_{lim}$ defined above). We consider that particles belong to these resonant regions as long as their periastron is larger than that of
the crossers, or equivalently, while they remain below the lower
dot-dashed green line of the figure.

Finally, we define region 1 as the region outside the resonances, and
below the limit of the crosser particles. Those regions are labeled
1a, 1xa, and 1xxa, for the particles below $e_{lim}$, and 1b, 1xb, and
1xxb for particles above $e_{lim}$.

From Fig. \ref{refsim} we observe that the giant planet alone cannot produce any significant perturbation on the disk, as expected, even after 1 Gyr. At the end of the simulation, only five particles have grown their $e_p$ above $e_{lim}$ through resonant perturbations at the 2:1 MMR of the giant. Not a single crosser is obtained for this case (neither at the end nor at any intermediate stage).

\section{Results and Discussion}
\label{results}

In this work we explore the evolution of cometary nuclei in a cold
debris disk under the gravitational influence of an interior giant
planet and a distribution of DPs. In this scenario resonances are the
most influential dynamical feature, but the dynamical evolution
resulting from the interaction with the DPs will lead some particles,
originally located away from the MMRs of the giant planet, into getting
trapped in such resonances; in fact we find that this combination (DPs
plus resonances) is much more influential than the resonances
alone (as can be seen from the reference simulation). In other words, we are interested on the dynamical paths that
replenish the giant's MMRs with cometary nuclei.

Due to computational constraints, we study the effect of a single
configuration of DPs along the disk. Other distributions will lead to
different individual fates of the particles in the disk, however, in a
statistical sense, we do not expect the rate of particle
resupplying to the MMRs to change significantly with different DP configurations. For this reason, we focus on the
exploration of the most important parameter, in this case, the total
mass of the DP distribution (equivalently, the mass of the debris
disk).

To test the validity of the previous statement, we have 
explored one case with a disk of 0.05M$_\oplus$ (using only 2000 test particles), where we have moved one of the two most massive DPs (the one closer to the 2:1 MMR; see Figures \ref{DPsavsr} and \ref{DPsavsm}) to an outer location on the disk. This is done for two reasons: one, it lets us explore the evolution of the disk with a significantly different spatial distribution of the mass, as we move the second largest DP, and two, it lets us test if the evolution and number statistics of cometary nuclei close and inside the 2:1 MMR is severely affected by this single massive planet. The results of
this case are shown in Section \ref{sec:largeDP}.

\begin{figure*}[htp]

  \centering
  \subfloat[``Zero-mass'' disk.]{\label{sfig:a}
  \includegraphics[width=.33\textwidth,height=4cm]{distM00.pdf}}
  \subfloat[Disk mass: 0.01M$_\oplus$]{\label{sfig:b}
  \includegraphics[width=.33\textwidth,height=4cm]{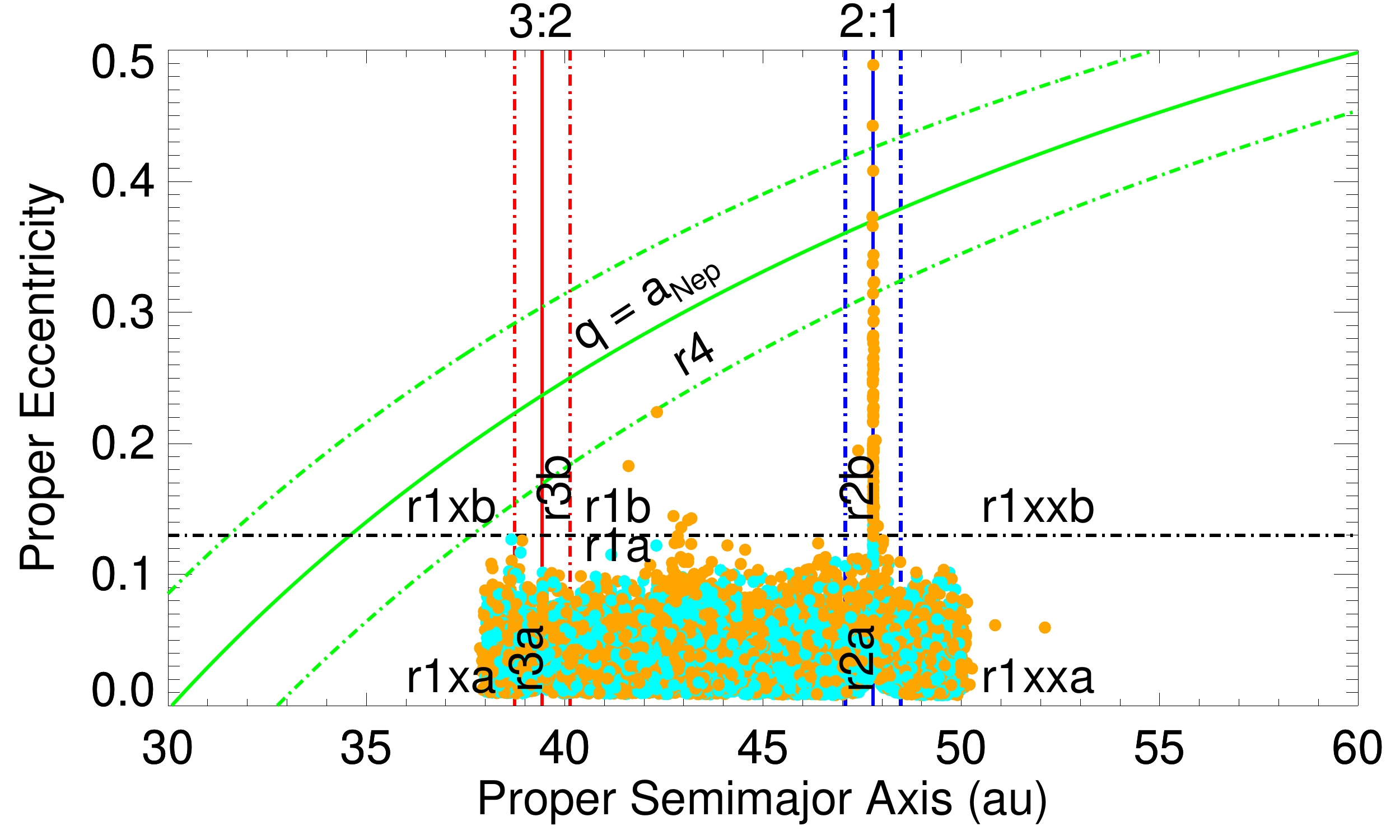}}
  \subfloat[Disk mass: 0.0223M$_\oplus$]{\label{sfig:c}
  \includegraphics[width=.33\textwidth,height=4cm]{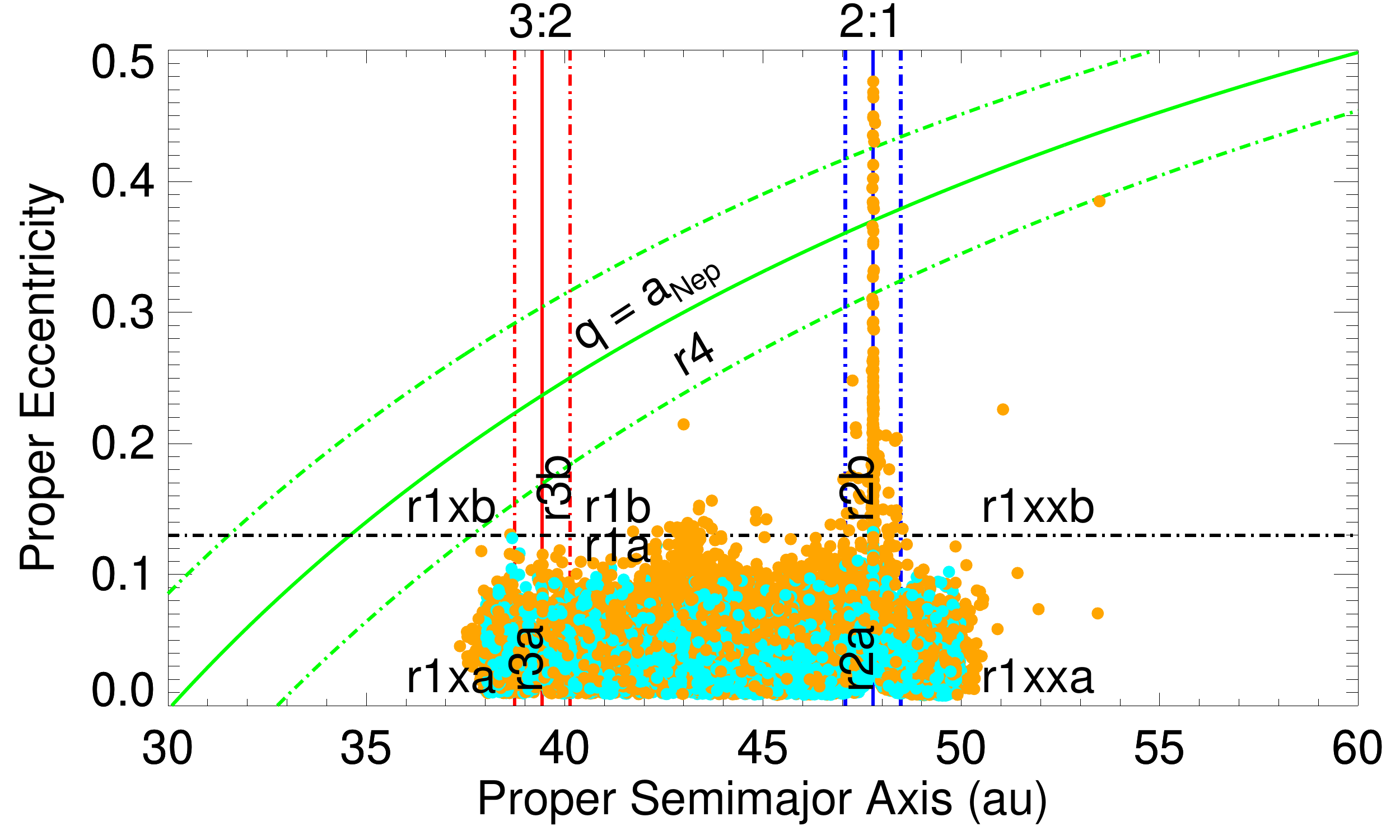}}
  \\
  \subfloat[Disk mass: 0.05M$_\oplus$]{\label{sfig:d}
  \includegraphics[width=.33\textwidth,height=4cm]{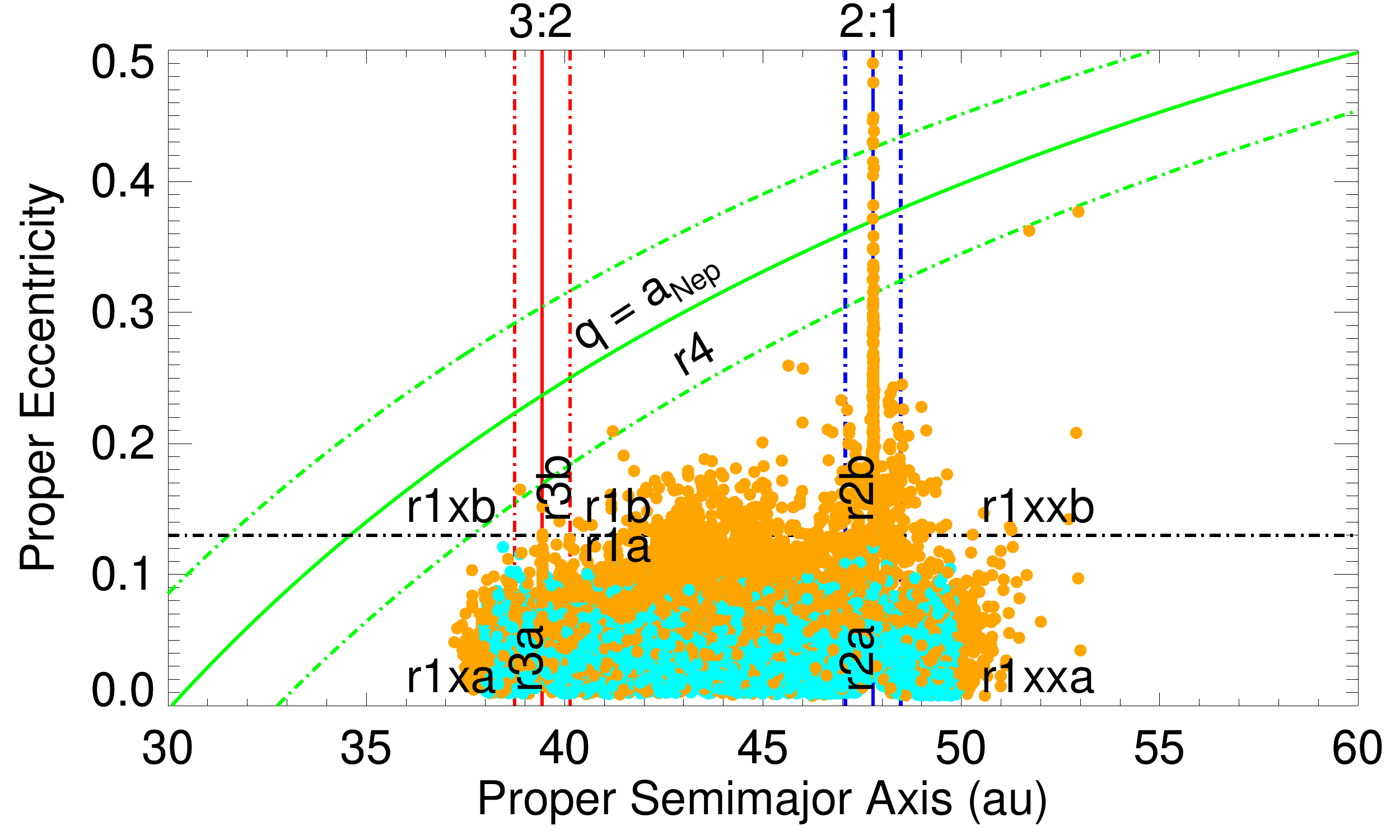}}
  \subfloat[Disk mass: 0.1118M$_\oplus$]{\label{sfig:e}
  \includegraphics[width=.33\textwidth,height=4cm]{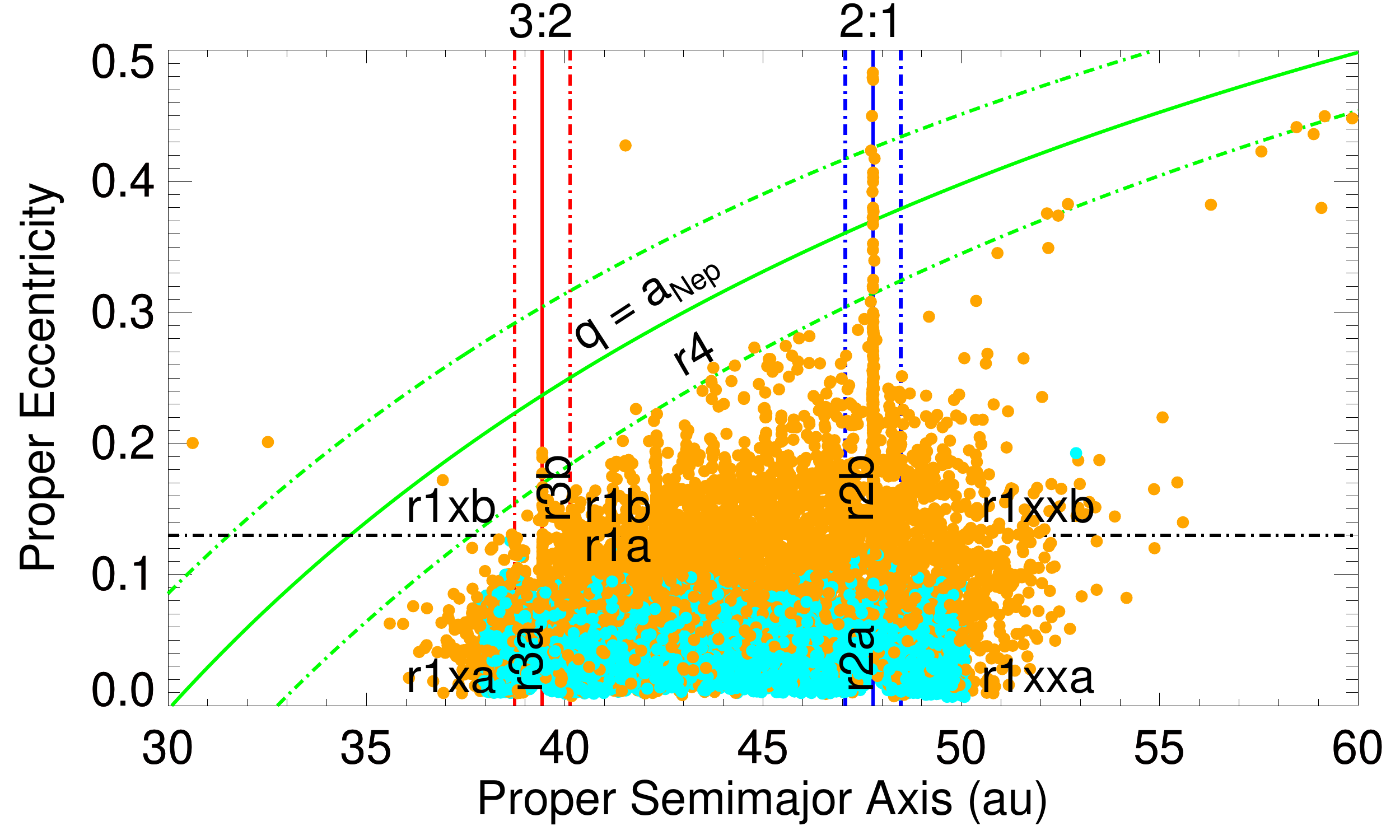}}
  \subfloat[Disk mass: 0.25M$_\oplus$]{\label{sfig:f}
  \includegraphics[width=.33\textwidth,height=4cm]{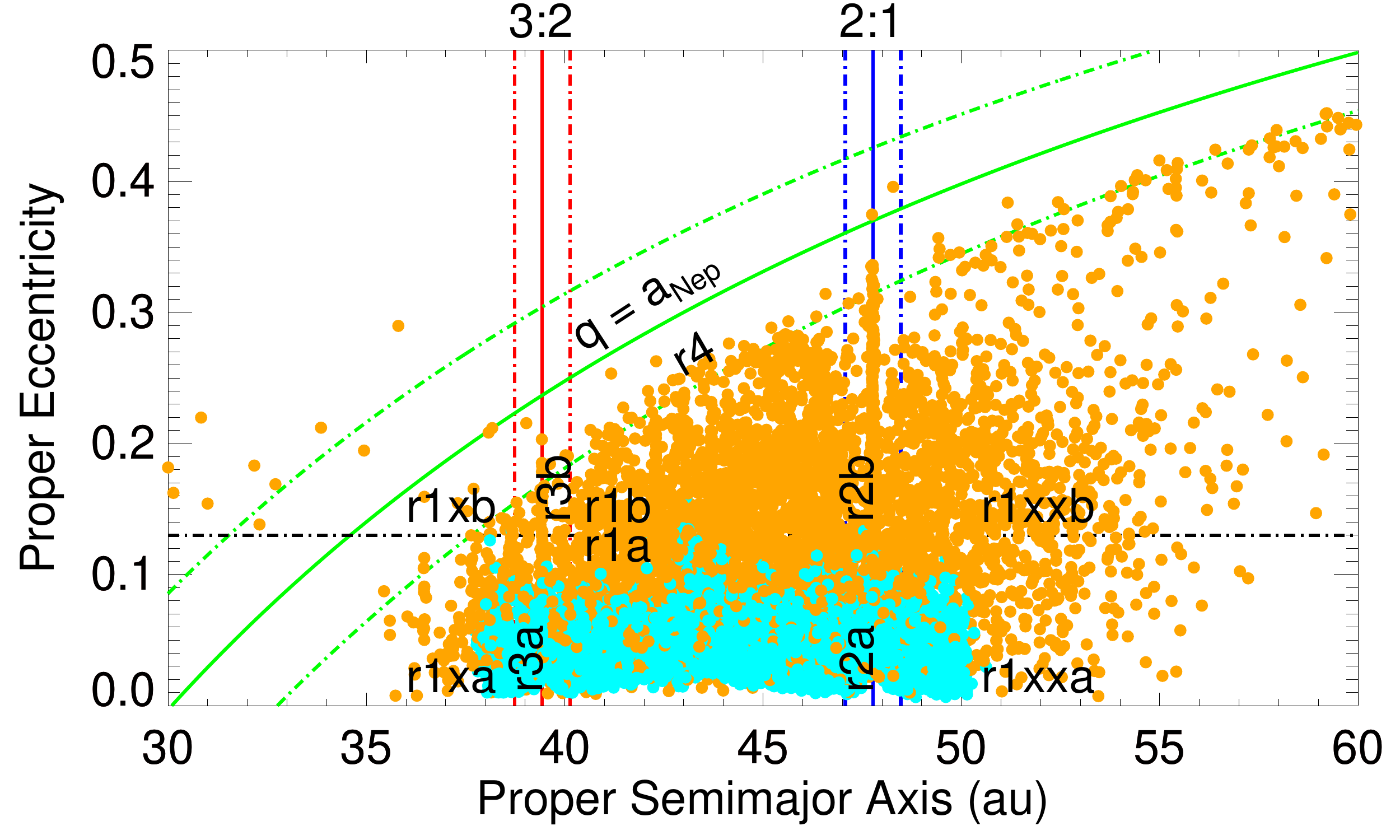}}

\caption{Initial and final point distributions of disk particles' proper semimajor axes and eccentricities, as a function of disk mass.\label{fig:distpae}}

\end{figure*}

\subsection{Evolution on the $a_p$ vs. $e_p$ plane}
\label{Evsep}

Fig. \ref{fig:distpae} shows the first sets of data points (cyan dots) and the last sets of data points (orange dots) of test particles in the plane $a_p$ vs. $e_p$, for all the six different total masses of the DP distributions considered in this study. From panel \ref{sfig:a}
to panel \ref{sfig:f} the masses are 0, 0.01, 0.0223, 0.05, 0.1118,
and 0.25M$_\oplus$. Note that the cyan dots do not represent the
orbital parameters at $t=0$ but rather they represent the average over the first ten Myr, hence the cyan distributions will not be identical in the six panels.

\begin{figure}
\plotone{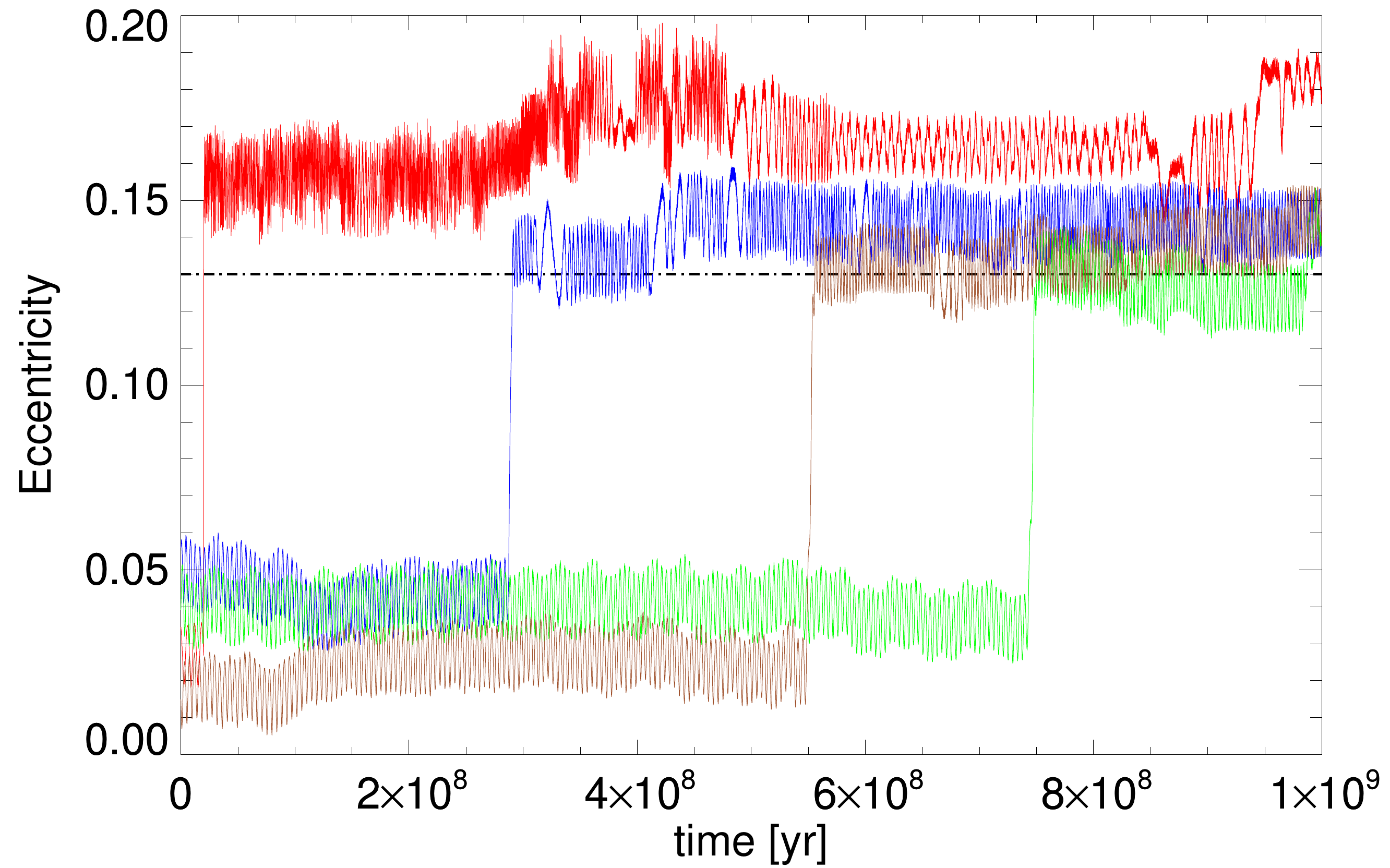}
\caption{Eccentricity evolution of four region 1 particles, which are perturbed and stirred by close encounters with DPs. The black dot-dashed line indicates $e_{lim}$.
\label{fig:nores}}
\end{figure}

A disk with a mass equal to 0.01M$_\oplus$, added in the form of 50 DPs, is enough to 
have an impact on the dynamics of the debris disk, as seen in
Fig. \ref{sfig:b}. The presence of a mass similar to that of the
smaller estimates of the mass of the CKB does not overheat the disk, however, we found
that almost 13 times more particles are stirred into region 2b; this
is even able to produce 19 crossers in 1 Gyr, as well as six
particles in region 1b, excited by close encounters with the DPs
alone. In order to illustrate this behavior, in Fig. \ref{fig:nores}
we plot the evolution of the eccentricity for four of the particles
initially located in region 1a which end up the simulation at region
1b. Those particles get their eccentricity stirred very quickly, in
some cases after a single very close encounter with a DP, e.g. the red
particle of Fig. \ref{fig:nores} suffered an encounter with a minimum
approximation distance, $d_{min}$, of just $9\,411$ km with a DP of mass
equal to $0.0022$ M$_\oplus$ (approximately the mass of Pluto); on the other hand, a series of close encounters, as close as $d_{min}\sim5.9\times10^5$ km, with several DPs in a short time-span, typically of the order of 1 to 5 Myr, can
quickly stir the eccentricities, as is the case for the blue, green,
and brown particles in Fig. \ref{fig:nores}.

\begin{figure}
\plotone{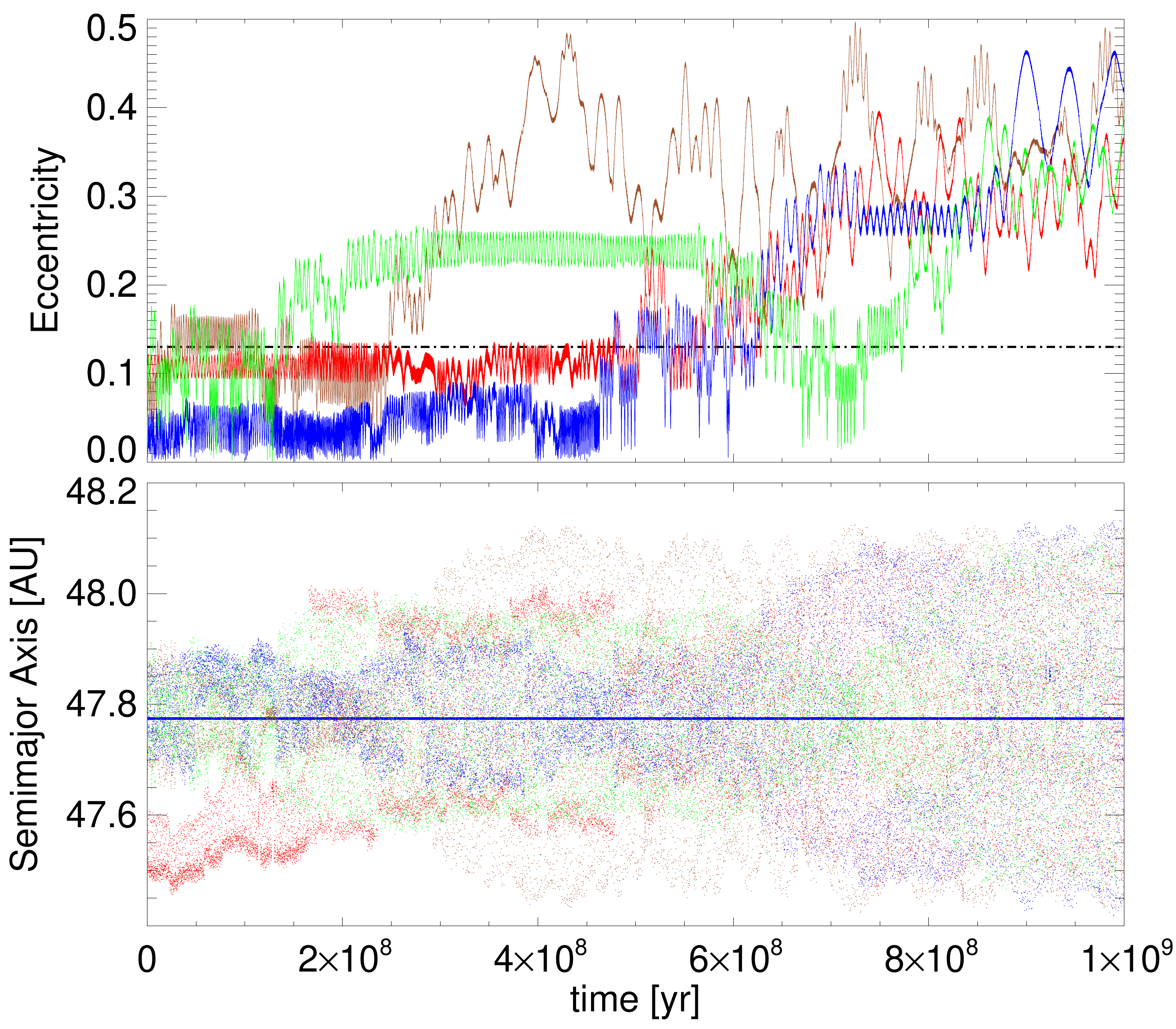}
\caption{Eccentricity (top) and semimajor axis (bottom) evolution of four 2:1 resonant particles that become crossers before the end of the 1 Gyr simulation. The black dot-dashed line in the upper panel indicates $e_{lim}$, while the solid blue line indicates the center of libration of the 2:1 MMR.
\label{fig:res}}
\end{figure}

Resonant particles increase their eccentricities by a completely
different mechanism. In Fig. \ref{fig:res} we show the evolution of
eccentricity and semimajor axis for four of the crosser particles that
reach region 4 through the 2:1 MMR with the giant. These particles
stir their eccentricities and inclinations chaotically \citep{Tiscareno09}, while maintaining a relatively bound semimajor axis. The semimajor axes
variations are related to the amplitude of libration of the
corresponding resonant argument. The larger those amplitudes, the less
tightly trapped in the resonance particles are, thus they are able to
escape more easily from the resonant region. From Fig. \ref{fig:res}
we can see how large eccentricities are correlated to large variations
of the semimajor axis of the resonant particles. The slow process that
leads to the growth of the orbital elements and consequently to the
escape of resonant particles is called chaotic diffusion.

In the trans-Neptunian region of the solar system, the slow chaotic
diffusion that lets particles evolve from MMRs into scattered disk objects, Centaurs, and short-period comets, has been studied
elsewhere \citep{Duncan95,Morbidelli97,Tiscareno09}. Previous studies have
focused on the slow chaotic diffusion originated by the overlapping of
secondary resonances, such as secular and Kozai resonances
\citep{Duncan95,Morbidelli95,Morbidelli97}. This phenomenon is independent of the presence or absence of large DPs close or even
trapped inside the MMRs, at least for small disk masses;
for example, \cite{Tiscareno09} show that the inclusion of Pluto, in
the 3:2 MMR with Neptune, changes the mean particle lifetime inside the resonance by only 3\% for 1 Gyr long integrations (when compared with a model without Pluto). In Section \ref{sec:largeDP} we address this result by exploring the presence or absence, close to the 2:1 MMR, of a DP larger in size than the Moon (for a disk of 0.05M$_\oplus$), finding
statistically equivalent results for the number
and evolution of resonant particles.

\begin{figure*}[htp]
  \centering
  \subfloat[``Zero-mass'' disk.]{\label{scros:a}
  \includegraphics[width=.33\textwidth,height=4cm]{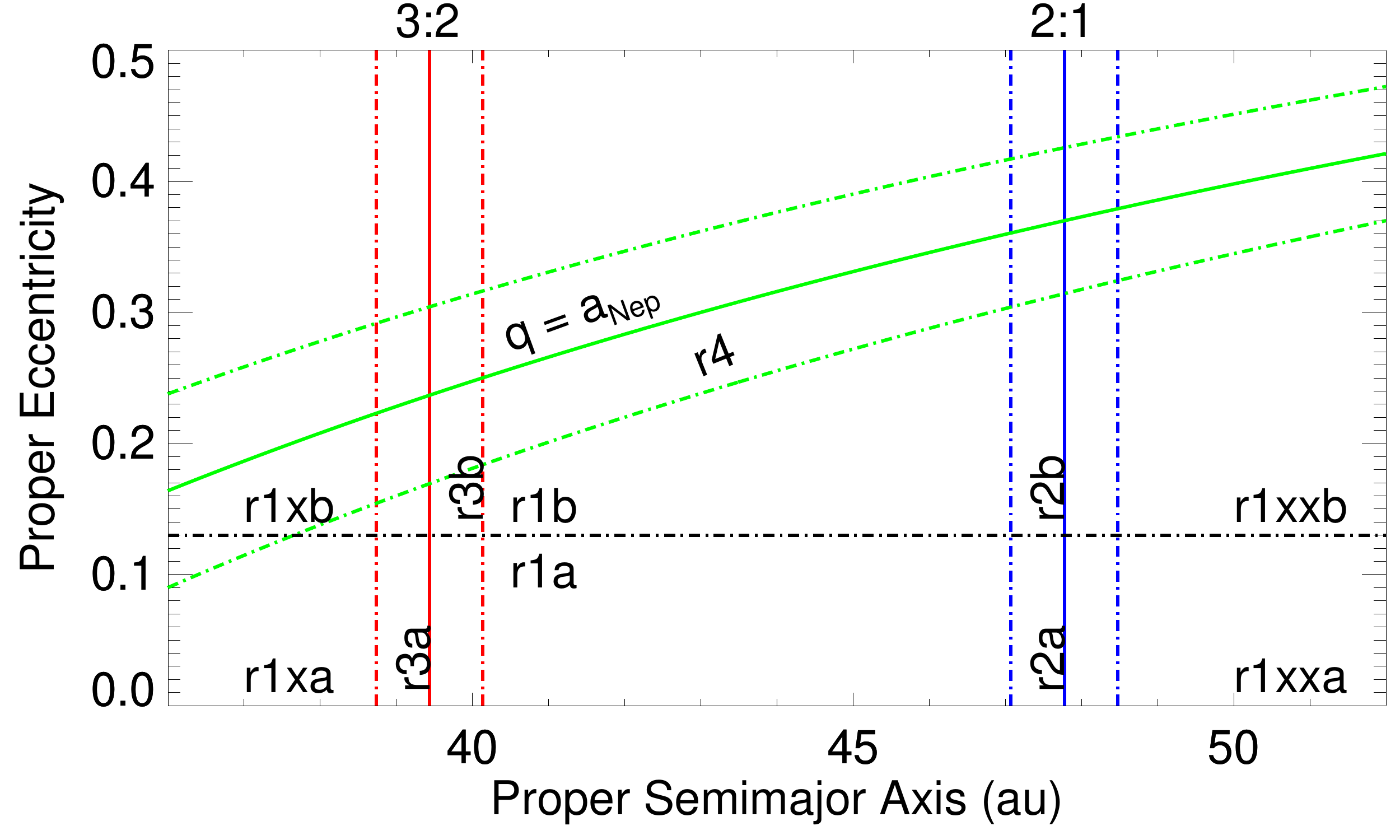}}
  \subfloat[Disk mass: 0.01M$_\oplus$]{\label{scros:b}
  \includegraphics[width=.33\textwidth,height=4cm]{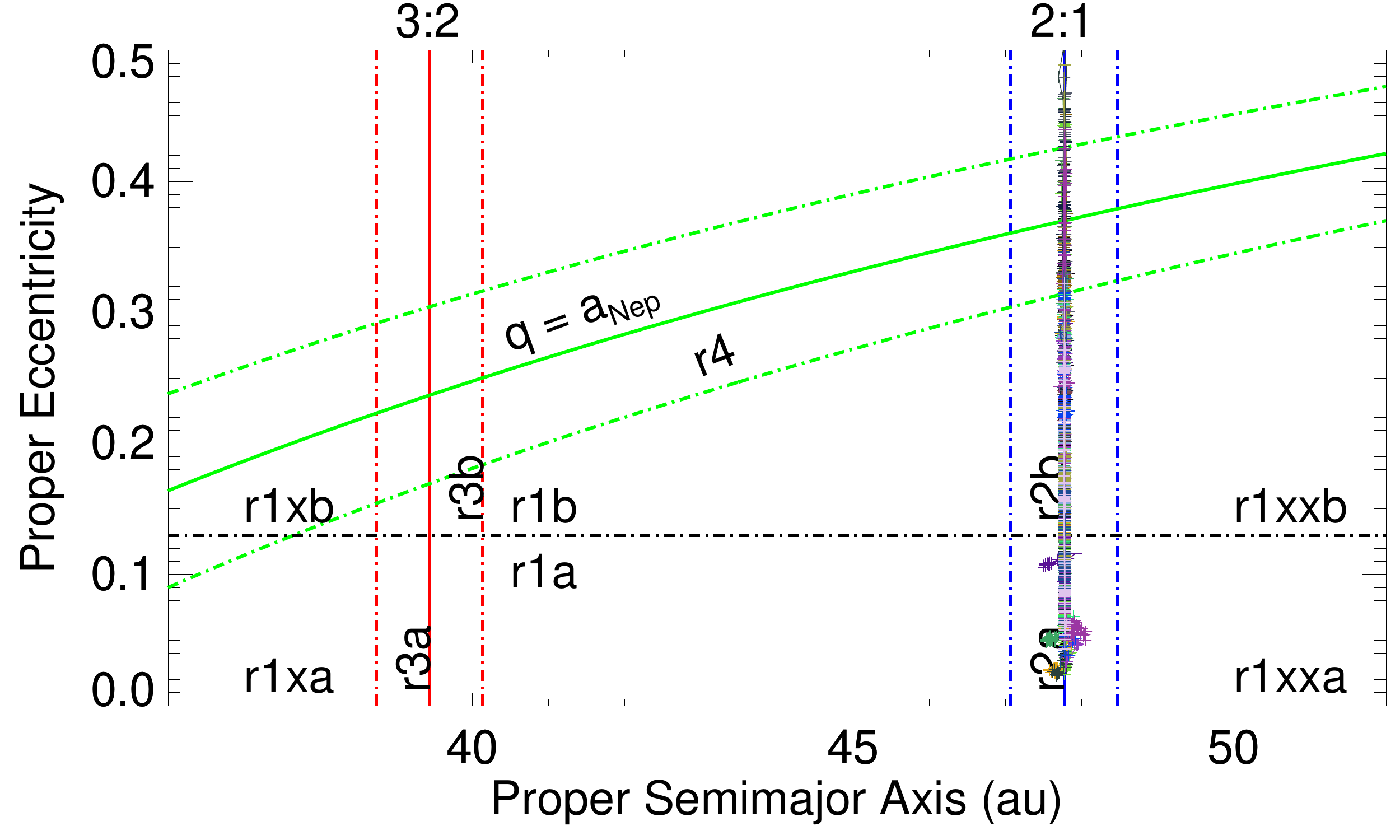}}
  \subfloat[Disk mass: 0.0223M$_\oplus$]{\label{scros:c}
  \includegraphics[width=.33\textwidth,height=4cm]{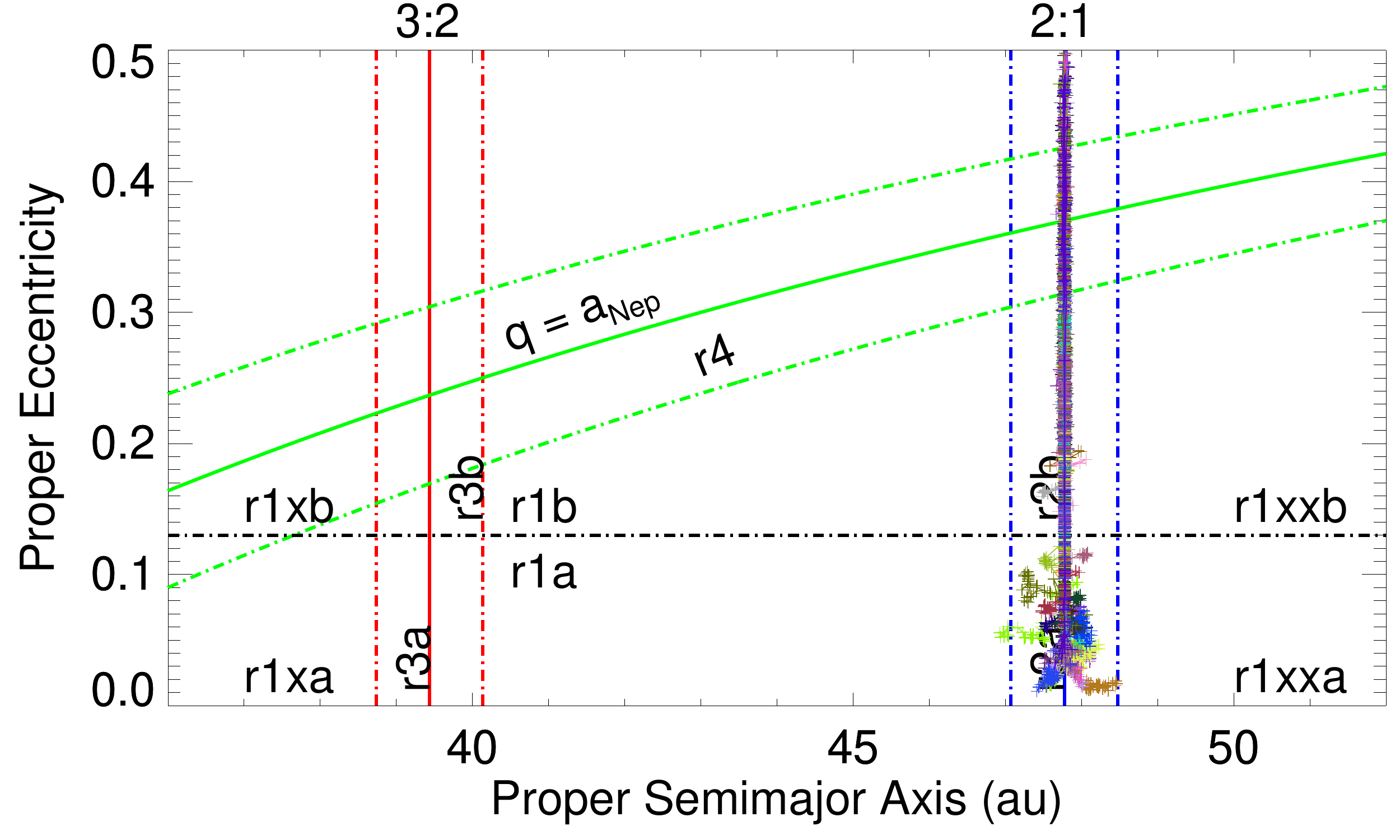}}
  \\
  \subfloat[Disk mass: 0.05M$_\oplus$]{\label{scros:d}
  \includegraphics[width=.33\textwidth,height=4cm]{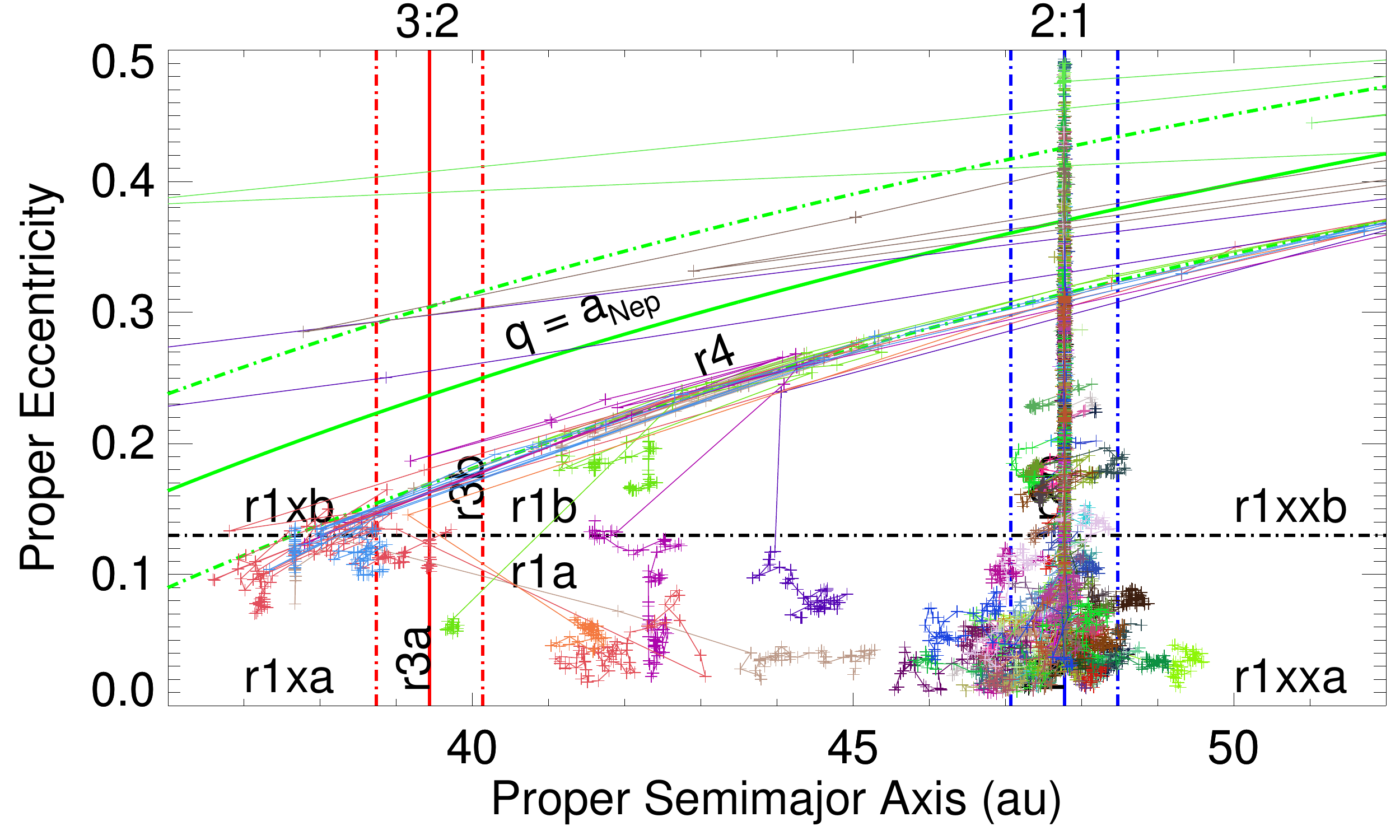}}
  \subfloat[Disk mass: 0.1118M$_\oplus$]{\label{scros:e}
  \includegraphics[width=.33\textwidth,height=4cm]{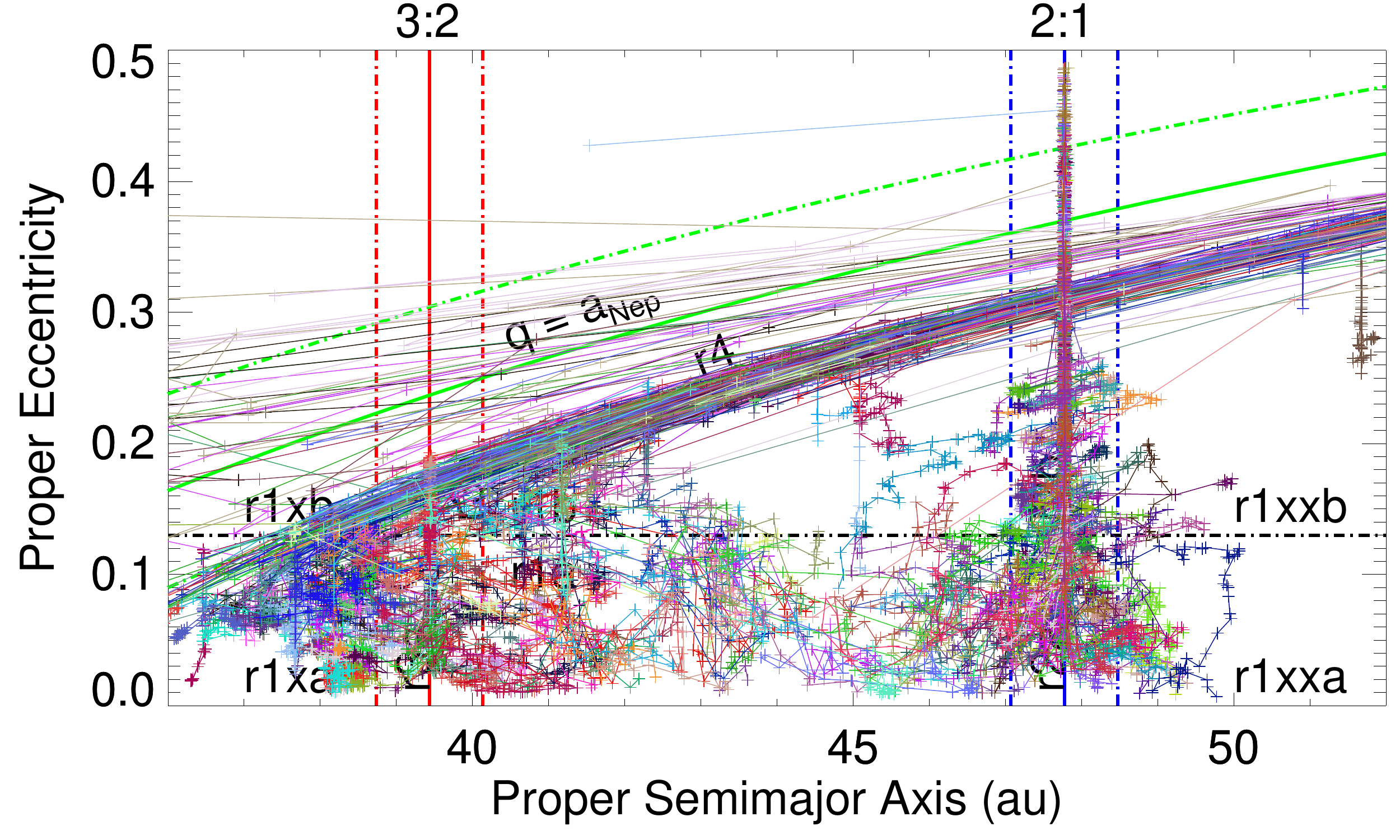}}
  \subfloat[Disk mass: 0.25M$_\oplus$]{\label{scros:f}
  \includegraphics[width=.33\textwidth,height=4cm]{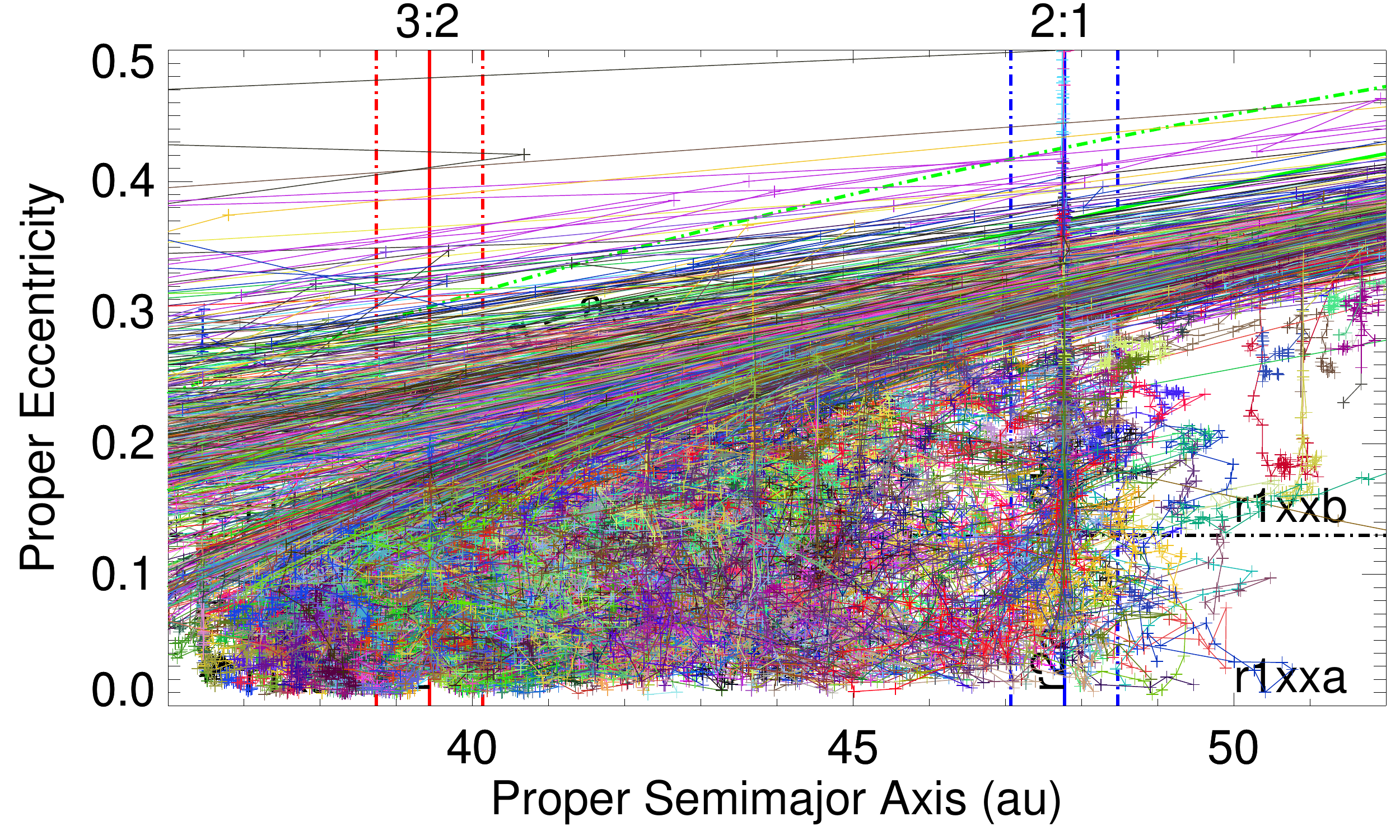}}

\caption{Track paths of all crosser particles in each simulation as a function of disk mass. The panels have 0, 19, 42, 70, 138, and 592 crossers respectively. The two most massive disks produce too many crossers to follow each particle individually. Each track that goes to the left of the panel represents a particle that visits the inner planetary system (at least briefly); each track that exits the panel to the right represents a particle that is scattered, either to the scattered disk or out of the system entirely. \label{fig:crossers}}

\end{figure*}

While increasing the mass of the disk, the global behavior of
particles is analogous to the previous 0.01M$_\oplus$ case, but it
grows monotonically as a function of the mass of the disk
\citep{Munoz17}, for example for masses of 0.0223 and 0.05 M$_\oplus$
(panels \ref{sfig:c} and \ref{sfig:d}), the crosser particles are
still overwhelmingly produced inside the 2:1 MMR. Actually, only two
non-resonant crossers are produced in the 0.05M$_\oplus$ disk (see
panel \ref{scros:d} ahead). On the other hand, a new family of
particles that approaches the giant planet, not through resonances,
appears; these particles are not protected by the resonant mechanism
and thus are prone to be affected by close encounters with the giant
planet and sent to larger semimajor axes while maintaining a short
periastron, giving origin to a scattered population, similar to the
scattered disk in the solar system \citep{Gladman08}.

When the mass of the disk equals 0.1118 and 0.25 M$_\oplus$ (panels
\ref{sfig:e} and \ref{sfig:f}) the effect of DP perturbations is more
drastic. Now several crossers can reach very small periastrons even
outside of resonances while the scattered population grows
significantly. Interestingly, the 2:1 resonant crosser population,
seems no longer protected from encounters with the giant when the mass
of the disk is equal to 0.25M$_\oplus$. This implies that a larger
fraction of the particles are sent into the scattered disk before they
can reach smaller periastrons, or larger eccentricities. The stronger
perturbations produced by more massive DPs increase the amplitude of
libration of resonant particles, making them less tightly bound to the
resonance, thus easily being ripped from it. This mechanism is analogous
to the one originally proposed by \cite{Ip97} as the escape pathway of
Plutinos in the solar system that contributes to the re-population of
short-period comets. Indeed, Ip and Fernandez focus on the evolution
of cometary nuclei inside Neptune's 3:2 MMR into short-period comets,
which were gravitationally scattered by hypothetical massive objects
(with diameter in the range 100-500 km) located outside the 3:2
resonance.

Regarding the particles in region 1, where most disk particles reside,
they are dynamically heated, with increasing eccentricity as a
function of disk mass. We have studied this effect in a previous paper
\citep{Munoz17}, as well as the stabilizing effect induced by the
giant planet below a threshold disk mass (a giant like Neptune, will
help stabilize the eccentricities of particles in the disk, if the
mass in DPs is less than 0.096M$_\oplus$); above this limit, the giant
contributes to the dispersion of the particles. This effect can be
seen as the rapid thickening of the disk in the last two panels of
Fig. \ref{fig:distpae}, which are above said threshold.


\begin{figure*}[htp]

  \centering
  \subfloat[Disk mass: 0.1118M$_\oplus$]{\label{scros10:e}
  \includegraphics[width=.49\textwidth]{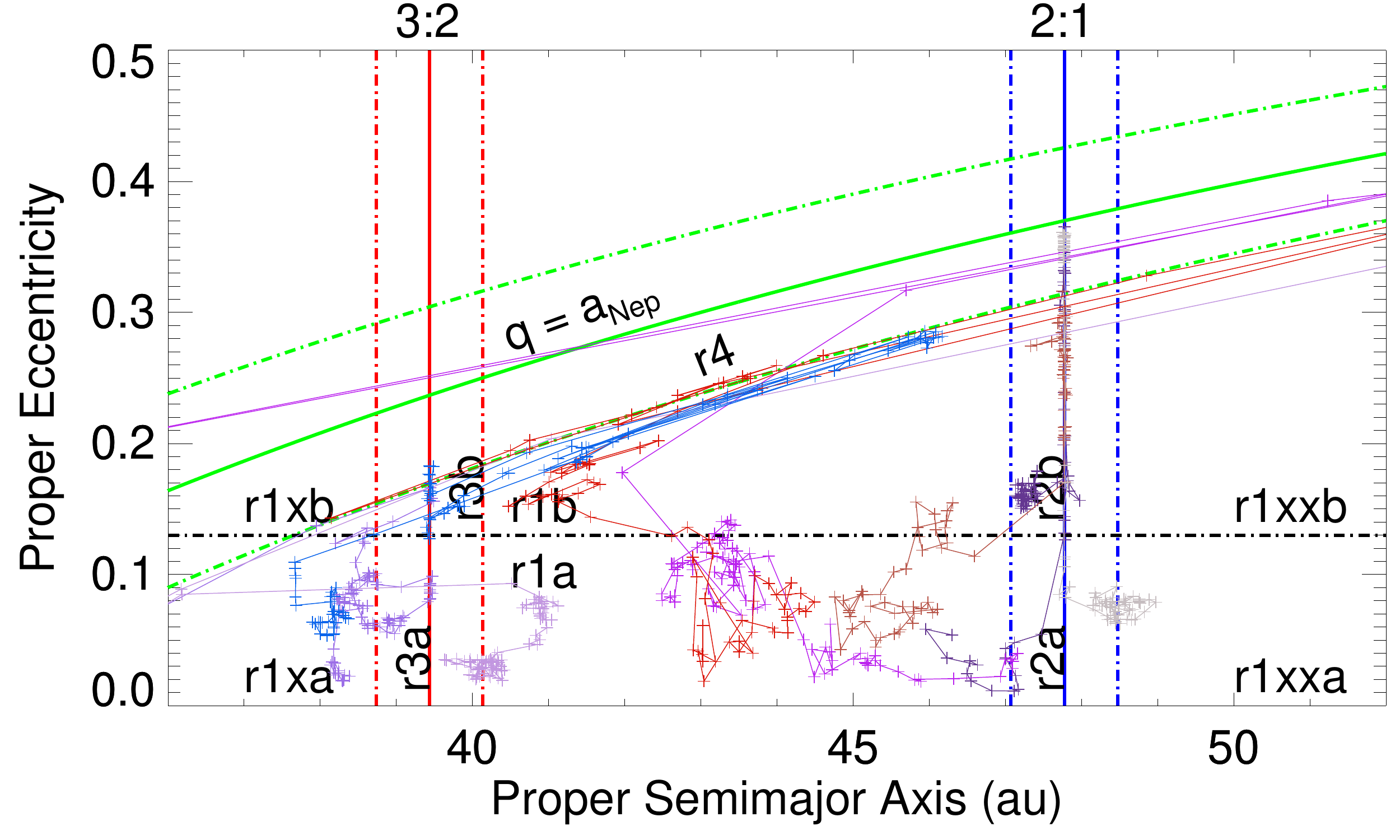}}
  \subfloat[Disk mass: 0.25M$_\oplus$]{\label{scros10:f}
  \includegraphics[width=.49\textwidth]{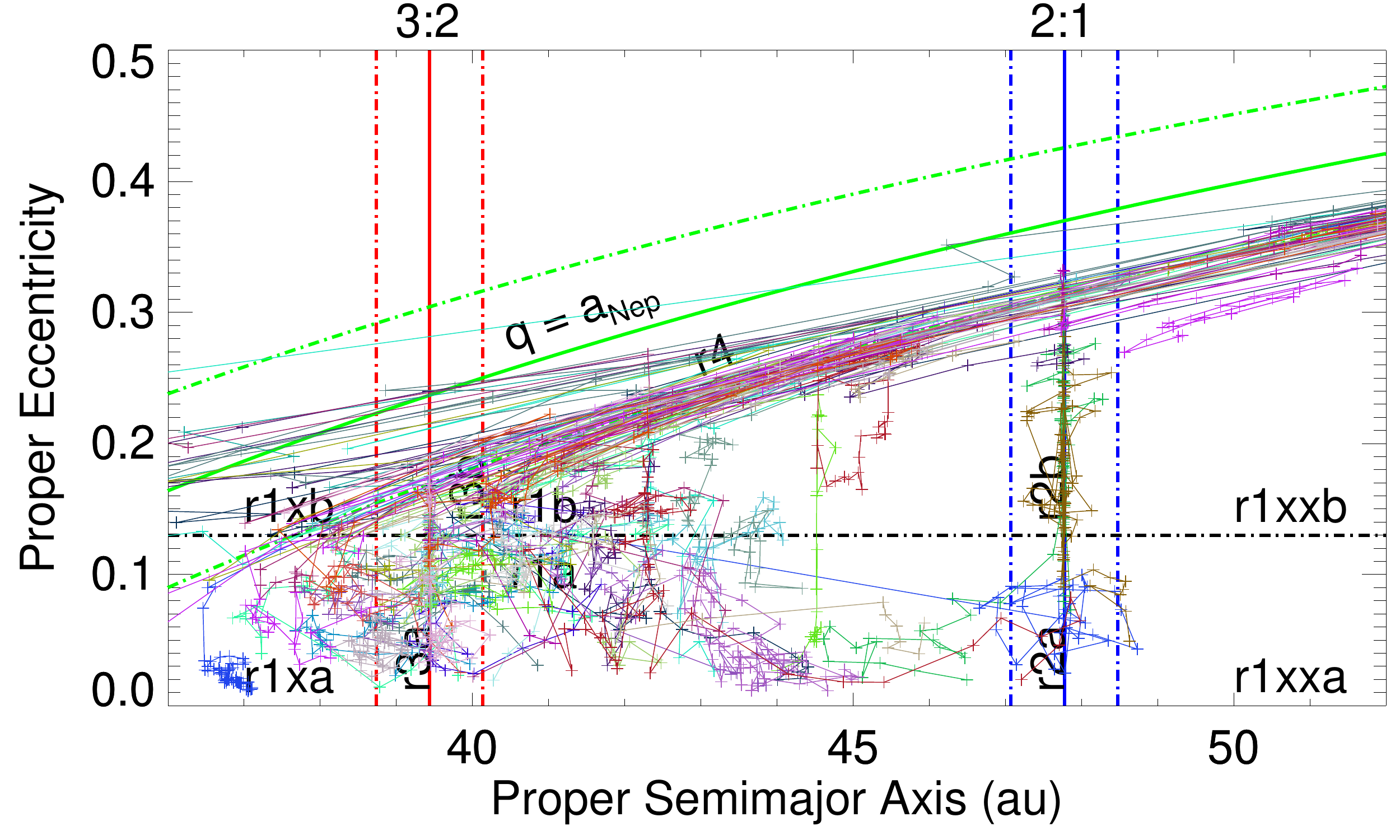}}

\caption{Same as panels \ref{scros:e} and \ref{scros:f} of Fig. \ref{fig:crossers}, with only $\sim$5\% of particle trajectories displayed for clarity. Left panel shows the track paths of 8 crossers while right panel shows 30 track paths.\label{fig:comp10}}

\end{figure*}

\subsection{Origin of the ``crosser'' population}
\label{crossers}

We are interested in the relation between the number and strength of
the perturbations suffered by resonant particles inside MMRs as a
function of the mass of the disk, or equivalently, as a function of
the mass of the perturbing DPs; particularly for those particles
within the 2:1 MMR. 

In this work we focus on the fraction of cometary nuclei evolving towards, and inside, the two stronger MMRs of the region. In Figure \ref{fig:crossers} we show the paths of crosser particles in the $a_p$ vs. $e_p$ plane, for all the disks masses explored. We include the zero mass panel for completeness (panel \ref{scros:a}), as no crosser is produced by the giant planet's perturbation alone.

As we did in Figure \ref{fig:distpae}, we increase the mass from panel
\ref{scros:b} to panel \ref{scros:f}, corresponding to 0.01 to 0.25
M$_\oplus$. From Figure \ref{fig:crossers} we can see how most of the
crossers come from the 2:1 MMR. Panel \ref{scros:b} corresponds to the lowest estimation for the mass of the CKB; here, all 19 observed crossers are produced
within the 2:1 MMR, and all of them come from the neighborhood of the
resonance. Interestingly, no crosser is produced from the 3:2 MMR,
even after 1 Gyr. This result cannot be compared directly to other solar system studies due to the homogeneous density of our simulation; in the case of the solar system, the main MMRs with Neptune, in particular the two stronger resonances, 2:1 and 3:2, become overpopulated during the outer migration of Neptune, with the consequent resonant sweeping of all the region interior to the present location of the resonances \citep[e.g.][]{Malhotra93}; thus, from the comparatively large number of resonant objects in the current Plutino population, one can expect a larger rate of escapees from this resonance, as observed in previous works \citep[e.g.][]{Morbidelli97, Tiscareno09}. It is interesting to note that, although the number of particles near and inside the 2:1 MMR in our work is essentially the same as in the 3:2 MMR (region 2a vs. region 3a, see also panel a of Figure \ref{fig:regfrac}), the effectiveness of the 2:1 resonance in producing crosser particles is much higher; this is consistent with the results for the solar system of \citet{Nesvorny00,Nesvorny01}, where particles with low eccentricity inside the 2:1 MMR with Neptune turn out to be unstable. In light of this result, even the relatively small population of twotinos in the
Kuiper belt should be of importance as a source of new scattered objects that could become short-period comets.

\begin{figure}
\plotone{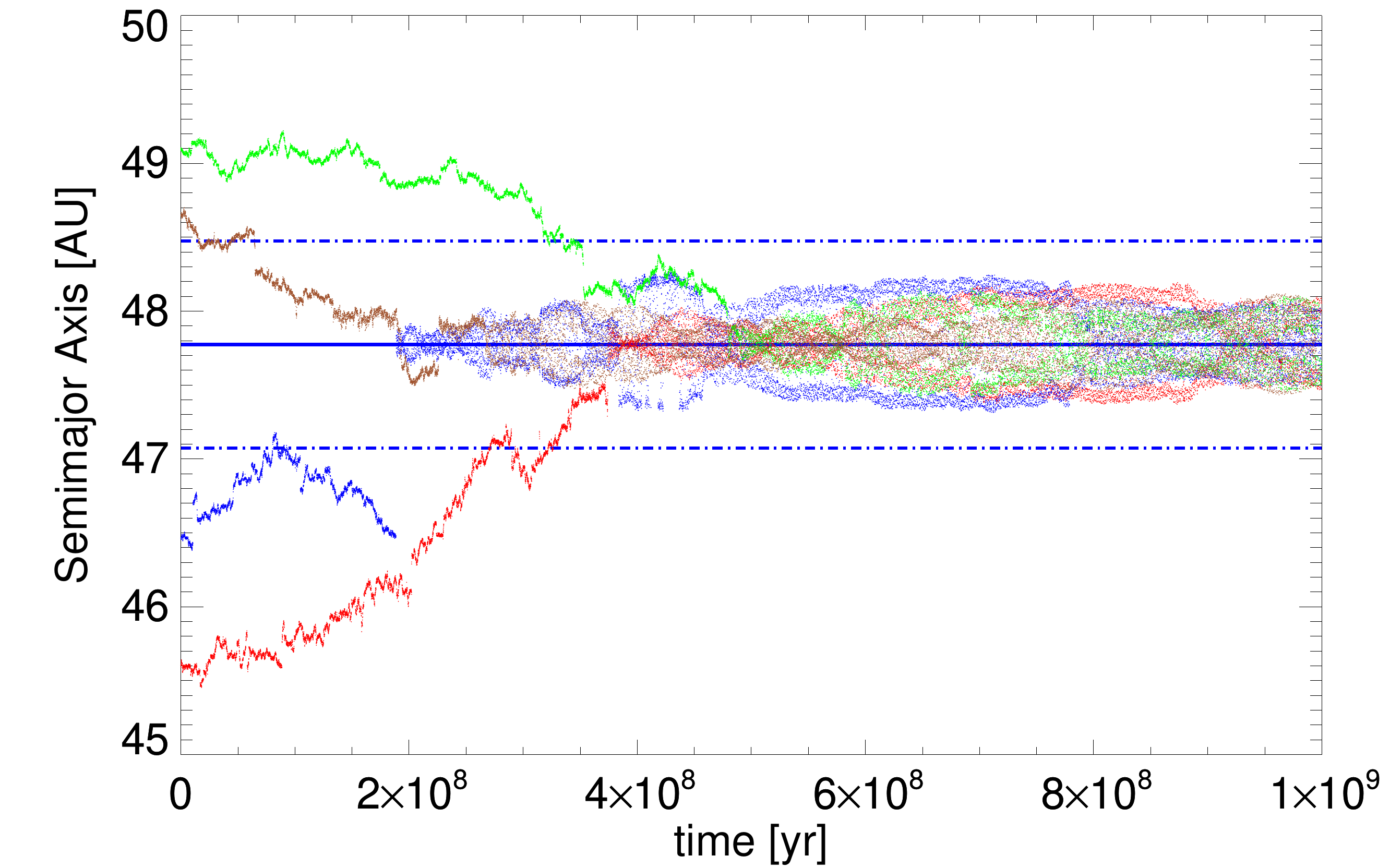}
\caption{Semimajor axis evolution of
four initially non-resonant particles that are sent to the 2:1 MMR region by perturbations from DPs. The solid blue line indicates the center of libration for the 2:1 MMR, while dot-dashed blue lines delimit our region 2.
\label{fig:resupp}}
\end{figure}

Panel \ref{scros:c} shows similar results than the previous one, with
a slight increase in the width, in the proper elements plane, of the
region from where crosser particles originate; this region is still inside the 0.7 au we used to define region 2; there is also an increase in the number of crosser particles from 19 to 42. In panel \ref{scros:d} we note an important new behavior: while still most crossers evolve through the 2:1 MMR, now six particles from region 1 (plus one from region 3a) are stirred by close encounters with DPs, overcoming the crossers limit and, by not being protected by a resonance, they are immediately and strongly perturbed by the giant, sending them to the scattered disk with large semimajor axis, or to the inner planetary system (see purple
particle originating from region 1a). Another crucial aspect, that
becomes evident in panel \ref{scros:d}, is the widening of the ``source
region'' of particles that evolve towards the 2:1 MMR. Now 
particles originally located even at $\sim$2 au away from the resonance
location, evolve into a resonant orbit, greatly increasing their
eccentricities, enough to become crossers. We further illustrate this behavior in Fig. \ref{fig:resupp}, where we show the semimajor axis evolution of four particles that started the simulation away from the 2:1 MMR libration center, which are sent to resonant orbits after some time. This mechanism, properly a resupplying of particles to the resonant region, is a direct product of the perturbations produced by massive DP objects, ranging in size, in this case (disk of mass equal to 0.05M$_\oplus$), from a bit larger than Ceres to slightly larger than the Moon.

Panels \ref{scros:e} and \ref{scros:f} correspond to disks with masses where the giant planet no longer acts as stabilizer \citep{Munoz17}, therefore the dispersion observed in those panels grows significantly when compared to the first four panels, with particles which come from region 1 being
rapidly sent either to the scattered population or to the inner planetary system system. In Fig. \ref{fig:comp10} we show a subset of crosser particle track paths (roughly 5\% of the total) for the two most massive disks, in order to illustrate with more clarity the evolution of the whole population of crossers for those cases. We again observe how the 2:1 resonant particles cannot reach the region of small periastron with moderate semimajor axis when the mass of the disk is 0.25M$_\oplus$, but instead they are quickly sent either to the scattered population or to the inner planetary system, as if they were not protected by the resonant mechanism at all.

\subsection{Fractional Evolution of the Populations}
\label{fracEV}

\begin{figure*}[htp]

  \centering
  \subfloat[``Zero-mass'' disk.]{\label{sfreg:a}
  \includegraphics[width=.33\textwidth,height=4cm]{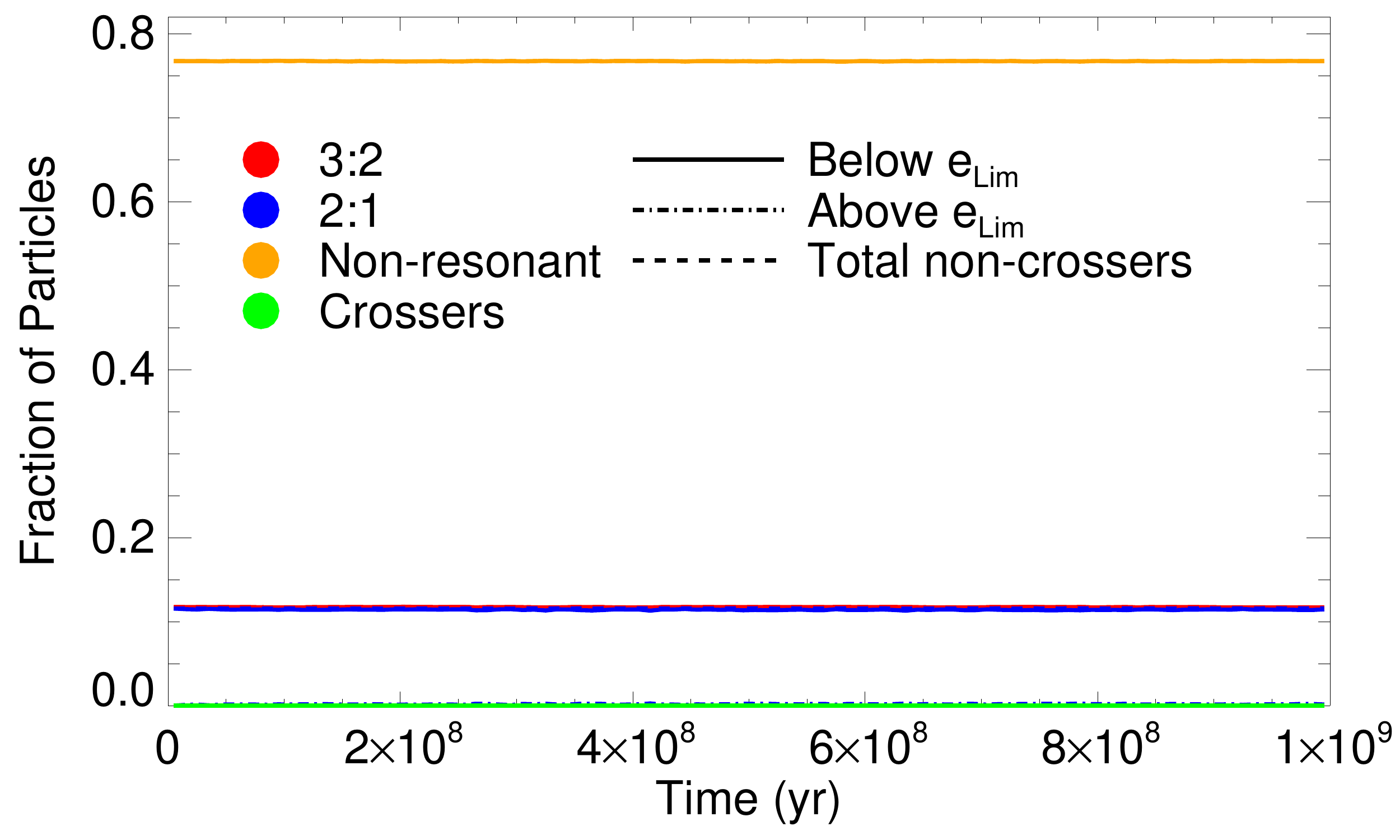}}
  \subfloat[Disk mass: 0.01M$_\oplus$]{\label{sfreg:b}
  \includegraphics[width=.33\textwidth,height=4cm]{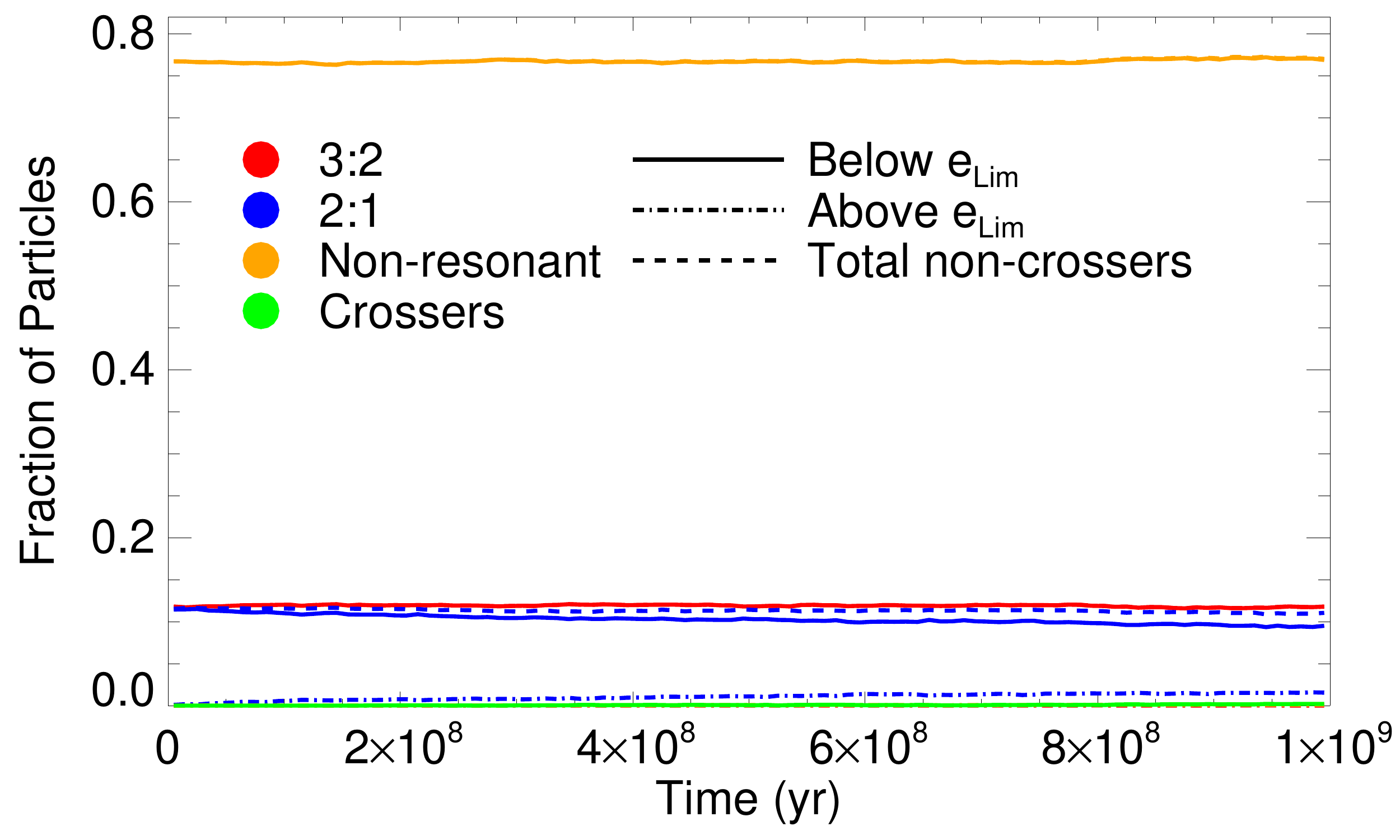}}
  \subfloat[Disk mass: 0.0223M$_\oplus$]{\label{sfreg:c}
  \includegraphics[width=.33\textwidth,height=4cm]{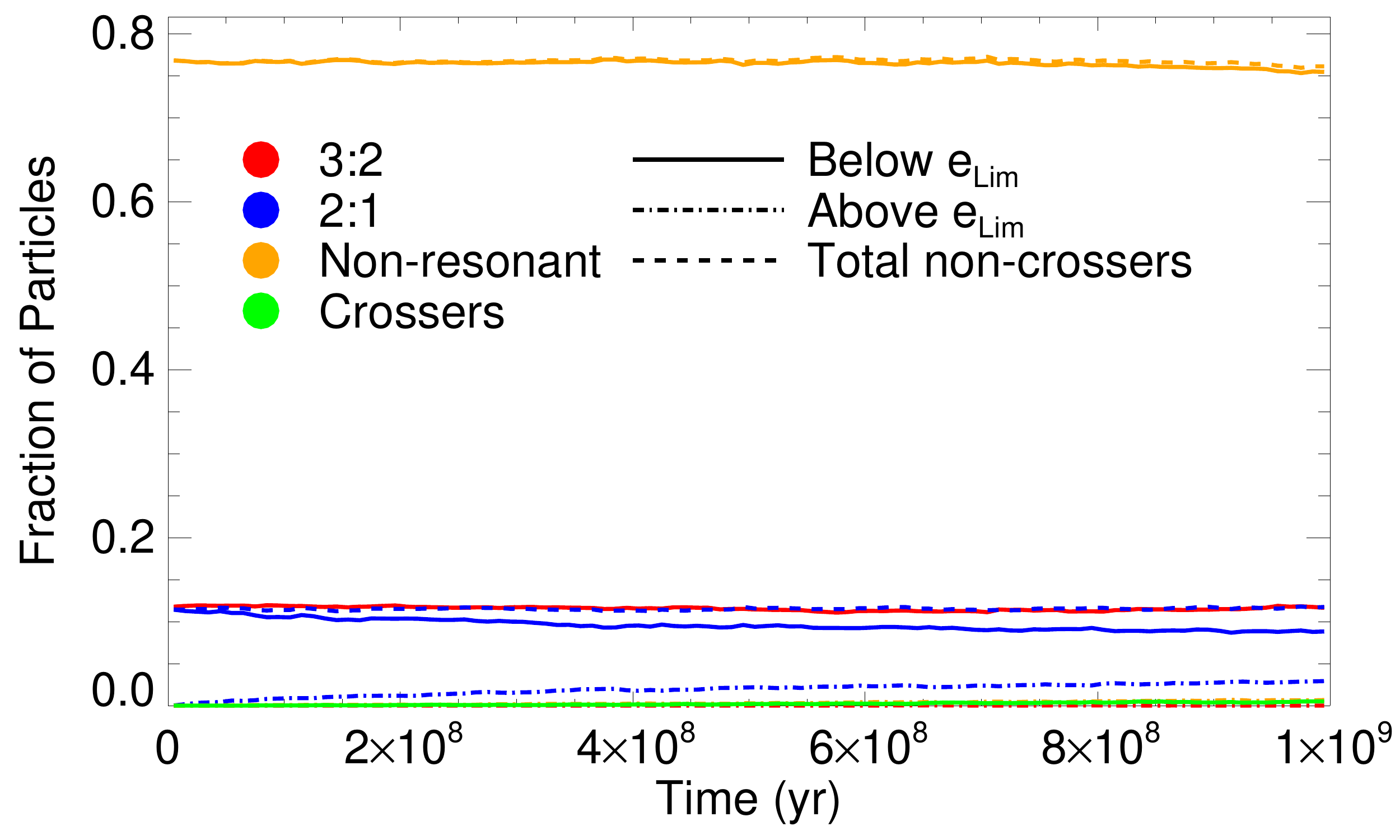}}
  \\
  \subfloat[Disk mass: 0.05M$_\oplus$]{\label{sfreg:d}
  \includegraphics[width=.33\textwidth,height=4cm]{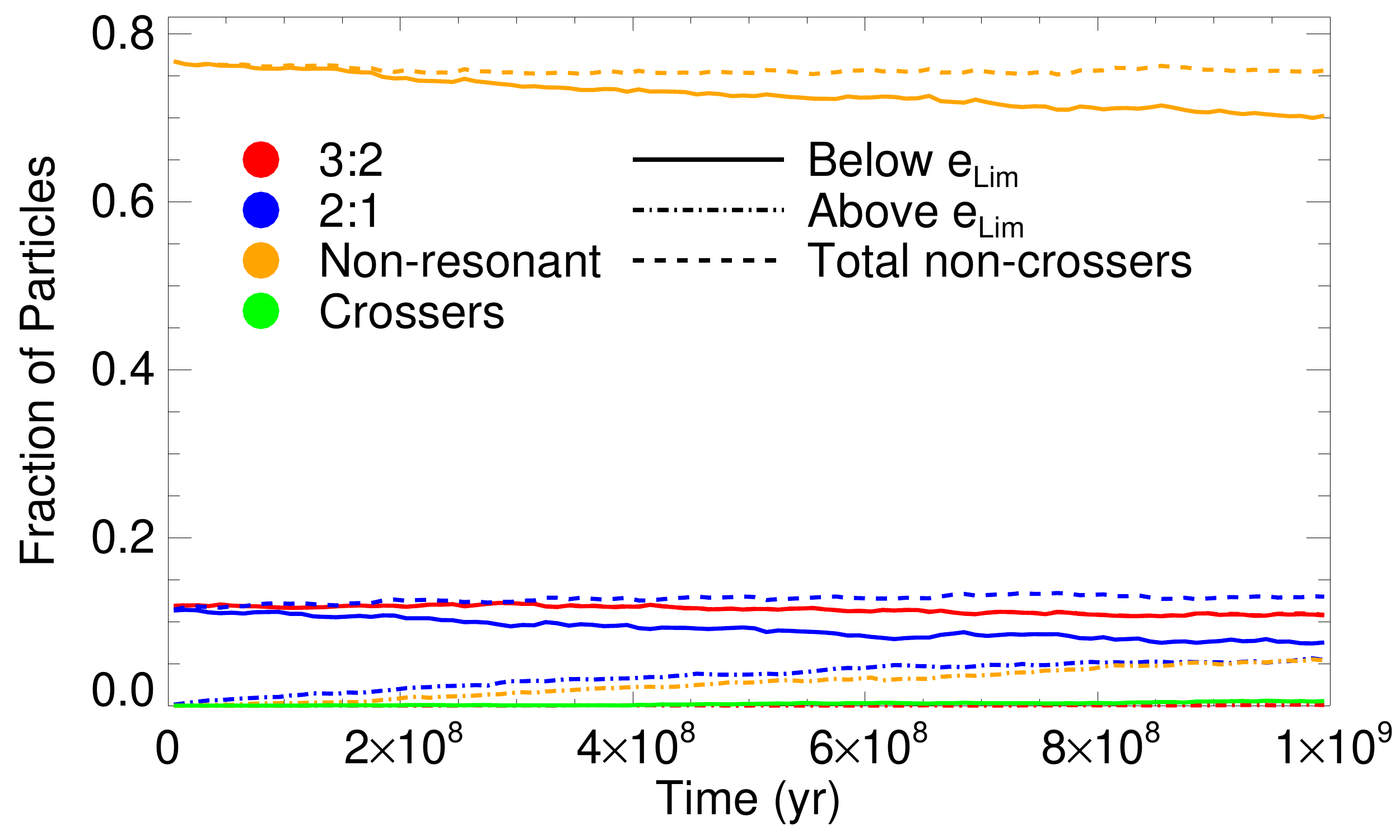}}
  \subfloat[Disk mass: 0.1118M$_\oplus$]{\label{sfreg:e}
  \includegraphics[width=.33\textwidth,height=4cm]{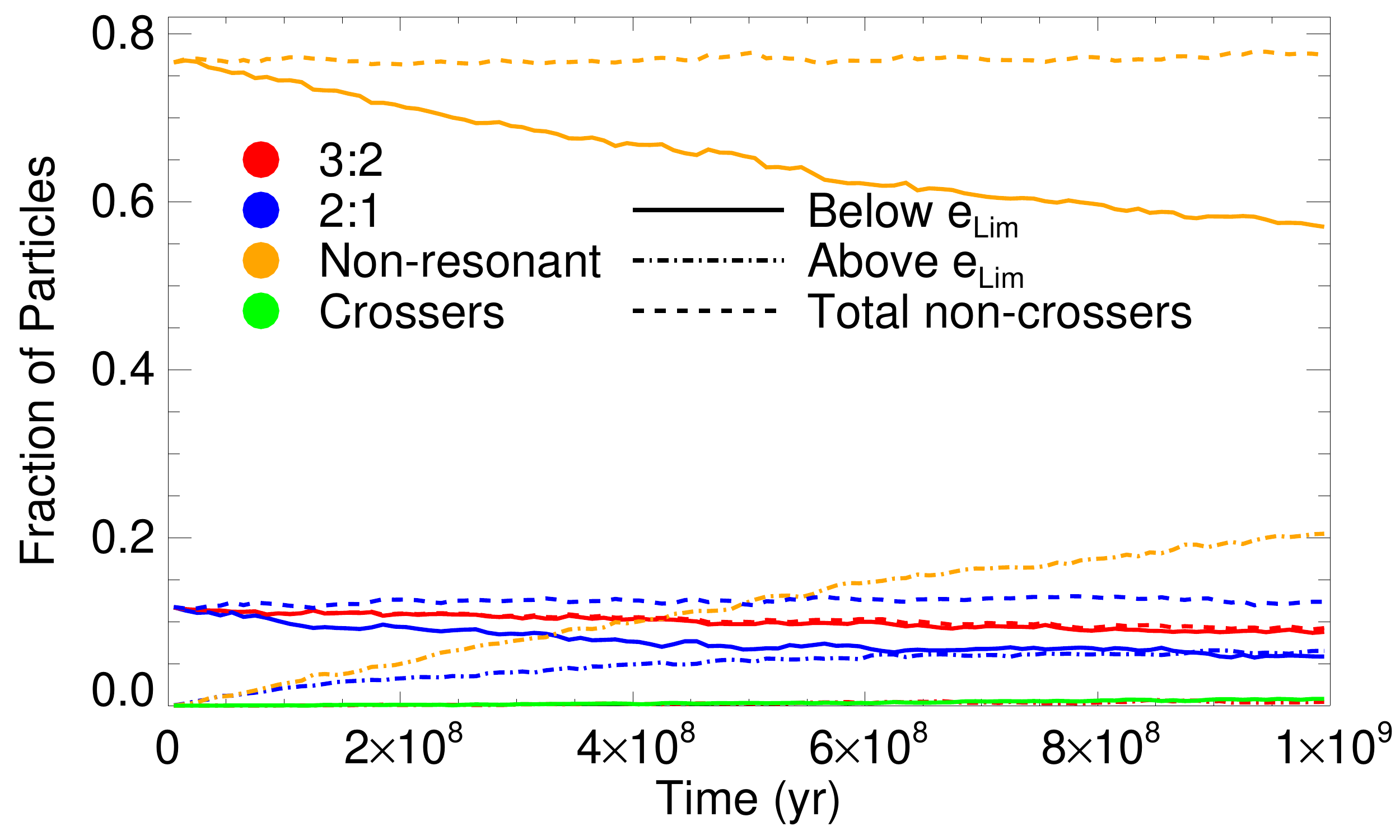}}
  \subfloat[Disk mass: 0.25M$_\oplus$]{\label{sfreg:f}
  \includegraphics[width=.33\textwidth,height=4cm]{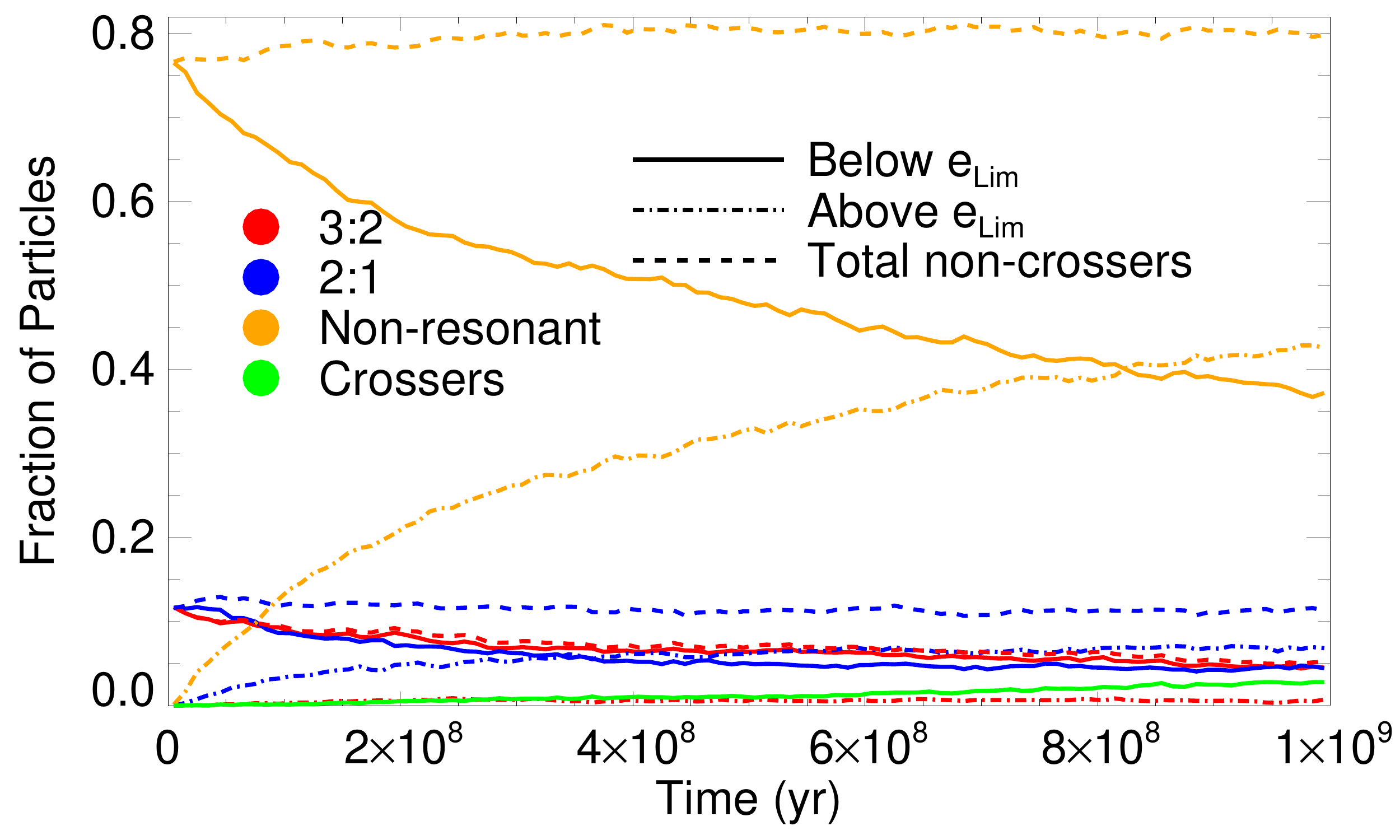}}

\caption{Fractional number evolution of the regions for all disk masses explored.\label{fig:regfrac}}

\end{figure*}

To quantify the diffusive effect of the DPs over the disk particle population,
we track the evolution of the fraction of particles in each of the 
four defined regions, during the 1 Gyr simulations. The results are shown 
in the six panels of Fig. \ref{fig:regfrac}. As in previous cases, from
panel \ref{sfreg:a} to panel \ref{sfreg:f} the mass 
of the disk increases from zero to 0.25M$_\oplus$.

The slope of the curves in all panels changes slowly with increasing
mass of the disk. Panel \ref{sfreg:a} shows a constant fraction of 
particles for all regions, as expected from previous results.
In panel \ref{sfreg:b} we notice a decrement in the fraction
of resonant particles below $e_{lim}$ (solid lines) in favor 
of resonant particles above $e_{lim}$ (dot-dashed lines), 
particularly for the 2:1 resonant particles (blue
curves). This trend continues in the following panels, while the 
total number of particles (dashed lines) in the resonant regions, 
both above and below $e_{lim}$, remains roughly constant for 
the 2:1 MMR; however, a slight 
increase in the total number of 2:1 resonant particles (dashed 
blue lines) is noticeable in
some panels of Figure \ref{fig:regfrac}. In contrast, the total 
number of 3:2
resonant particles (dashed red-line) closely follows that of the
3:2 resonant particles below $e_{lim}$ (solid red line), while 
the number of 3:2 
resonant particles above $e_{lim}$ (dot-dashed red line) does 
not grow 
significantly in any panel. This means that particles excited above
$e_{lim}$ through the 3:2 MMR are quickly scattered and are consequently lost to other families, 
while the whole population in this resonance gradually decreases.

The 2:1 total population is more or less
constant during all the integration, but given that several 
particles are passed from the 2:1 resonance region above $e_{lim}$ 
to the crosser population (solid green line), we can argue that an
effective resupplying mechanism is operating in favor of the 
2:1 MMR region, in order to maintain a steadily growing number 
of total region 2 particles while the crossers increase. 
Naturally, particles that resupply region 2 come from the
neighborhood of the resonance (i.e. region 1; orange lines). Therefore, part of the decrease in the number of total region 1 particles (or non-resonant 
particles) occurs due to the resupplying of the 2:1 MMR. This trend 
is maintained below the stabilizing limit of the giant, this 
is, a decrease in the total number of non-resonant particles 
(dashed orange lines) is maintained until panel \ref{sfreg:d}, 
but in the last two panels we observe that the total number of non-resonant particles increases slightly. Therefore, above the stabilizing 
limit of the giant, a great number of particles from region 1a 
(solid orange lines) are excited towards region 
1b (dot-dashed orange lines), while several others are sent 
towards the crosser region where they are scattered 
by the giant. 

Regions 1, 2, and 4 increase their numbers with the increasing mass of the disk, while the only family that suffers a decrement in the number of particles is region 3, those of the 3:2 MMR. Even if those particles cannot reach the 
crosser region through the 3:2 MMR, the majority of the 
perturbed particles coming from region 3 reach the scattered
part of the disk, where they are counted as part of the 1b family. Finally,
most of the crossers come originally from the 2:1 MMR region, 
with the second most abundant population of crossers coming from the 1a region (through the 1b region).

\subsection{The Crosser Production Efficiency of the 2:1 MMR is not a Result of a Single Large DP.}
\label{sec:largeDP}

\begin{figure*}[htp]

  \centering
  \subfloat[Fraction comparison.]{\label{scomp:a}
  \includegraphics[width=.49\textwidth]{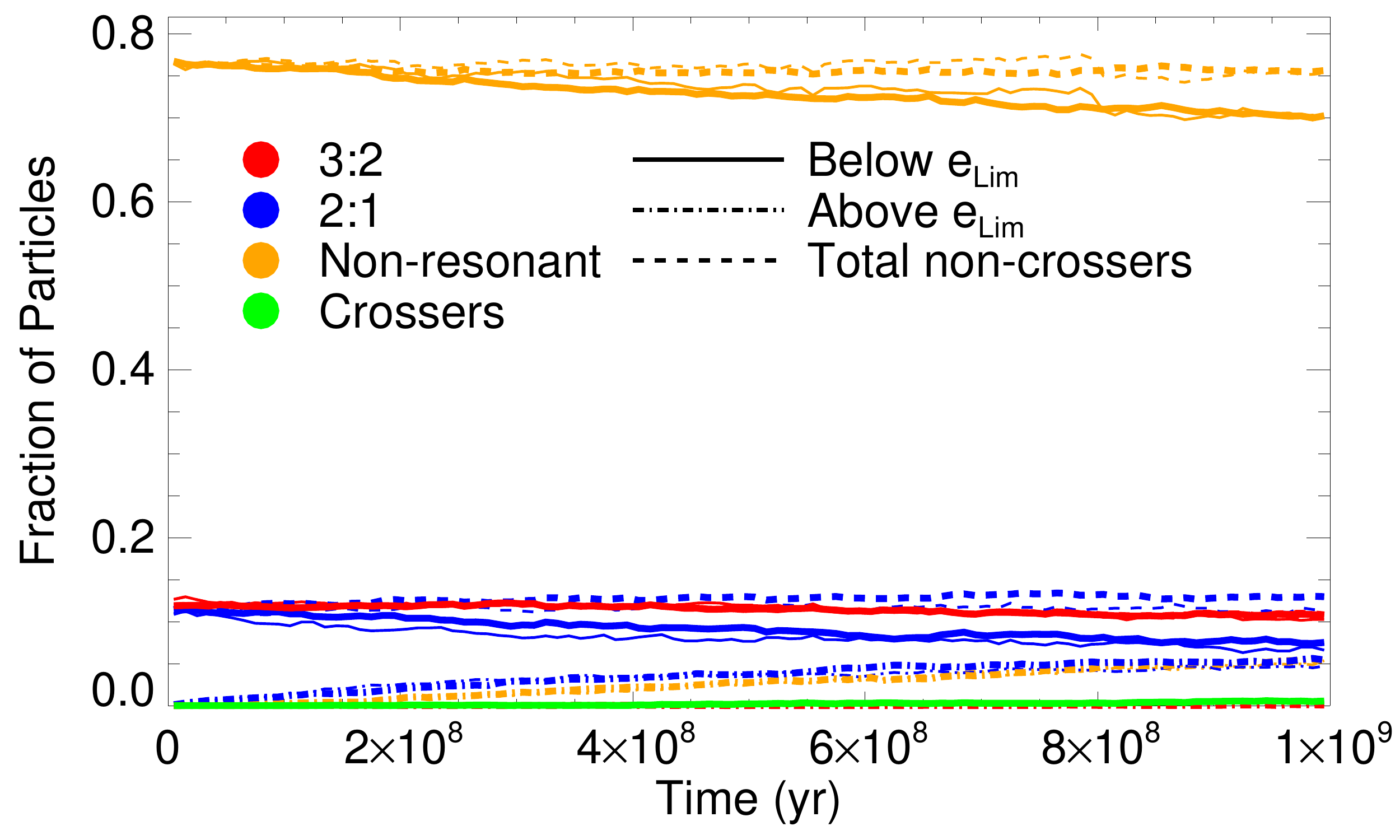}}
  \subfloat[Disk mass: 0.05M$_\oplus$]{\label{scomp:b}
  \includegraphics[width=.49\textwidth]{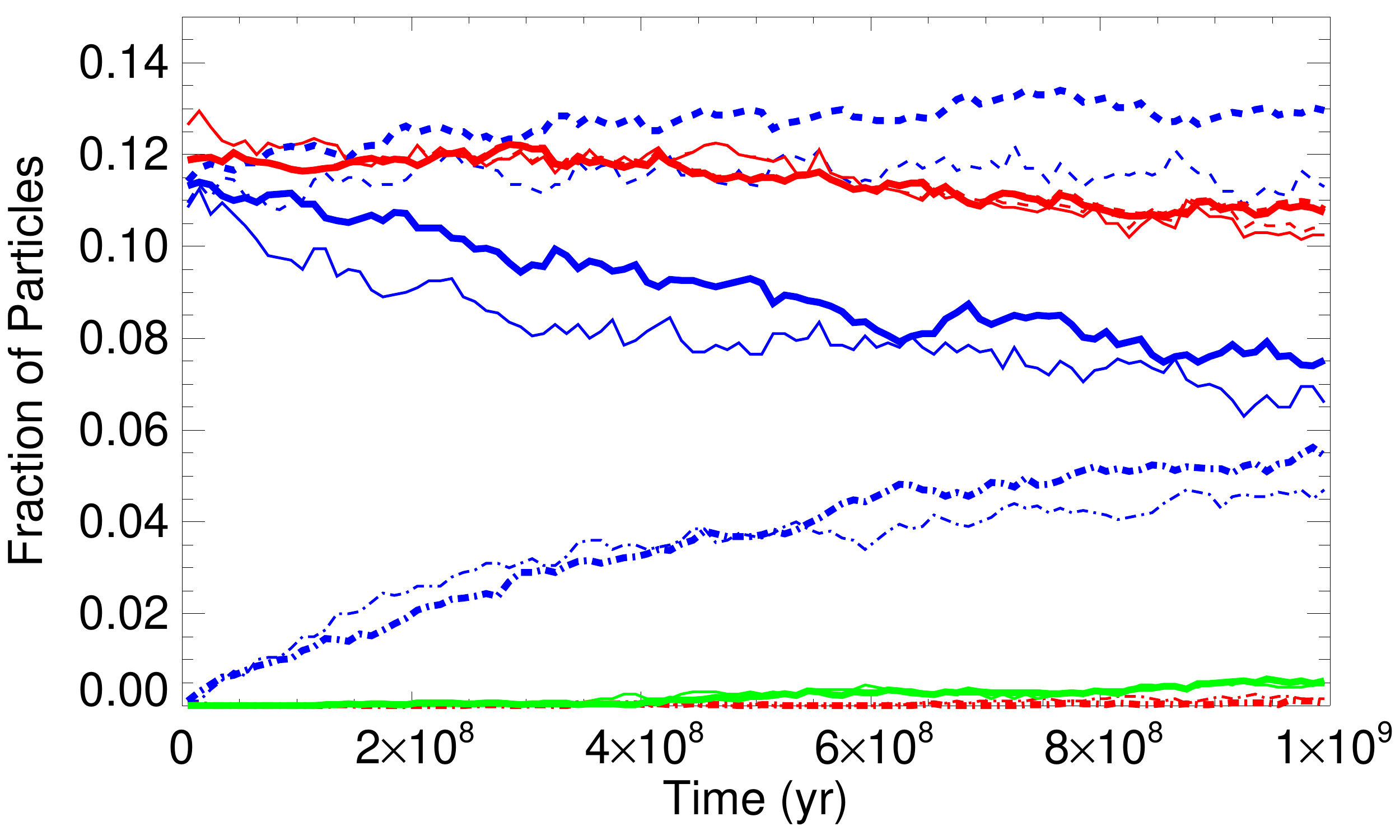}}

\caption{Comparison of the fractional number evolution for the simulations including 5000 (broad curves) and 2000 (thin curves) particles in a disk of 0.05M$_\oplus$. Panel a shows the full range coverage as Figure \ref{fig:regfrac}, while panel b shows an amplified range in order to compare regions 2,3, and 4.\label{fig:comp}}

\end{figure*}

The relevance of these results would be greatly diminished if they changed dramatically with a different distribution of DPs. In this section we present results from a single smaller simulation of a disk with mass equal to 
0.05 M$_\oplus$, where we have exchanged the second largest DP of the
distribution (the one closer to the 2:1 MMR in the rest 
of the simulations presented in this work) with one in the outskirts of the disk (beyond 50 au). All of the remaining characteristics of this
simulation are the same as those described in Section \ref{sec:sims}. We will show that this simulation turns out to be statistically equivalent to our standard model.

In Figure \ref{fig:comp} we compare the results of both of the 0.05M$_\oplus$ disk simulations. Our new simulation was done with 2000 particles; in \cite{Munoz17} we checked that simulations with that number of particles are statistically equivalent. Panel \ref{scomp:a} shows the fractional number of particles evolution for all the regions described earlier, where the broad curves represent results from our standard simulation, while the thin curves are result from the modified simulation. The fractional number evolution in all regions is visually almost indistinguishable from one another, but with slightly larger fluctuations in the new, smaller, simulation.

Not surprisingly, regions 1 and 3 do not suffer any significant variation due to the position change of the large DP. We expect a larger difference close to the region previously affected by this object. In panel \ref{scomp:b} we show a smaller region of the same plot shown in panel \ref{scomp:a}, where only the fractional number evolution of regions 2, 3, and 4 are visible. In this scale, we observe that the fractional number of crossers is the same also for region 4 in both simulations. However, some differences for region 2 become evident in panel \ref{scomp:b}. 

Indeed, the broad blue curves (region 2 of the standard simulation) are, on average, slightly above the thin blue curves (region 2 in the modified simulation), however at the end of the simulations, the large DP close to region 2 has been able to increase by $\sim$13\% the fractional number of particles belonging to the 2:1 resonant region. Overall, a very small change for such extreme modification. This shows that the large quantity of objects that reach region 2b is more closely related to the behavior of the 2:1 resonance and not to the specific population of DPs in the initial seeding of our model.

Regarding the crossers, the influence of the large DP seems to be even smaller. Based on the previous results, we would not expect significant changes for other ``statistically equivalent'' DP configurations.

\subsection{Injection Rate of Ecliptic Comets in a Solar System Toy Model.}
\label{toymodel}

As a first specific example, we will use our toy model to try to determine if the mechanism introduced here (the injection of ecliptic comets due to the replenishment of giant's MMRs by DPs) is relevant in the solar system. This is a preliminary estimation, a more detailed model is needed to obtain accurate results; we will study this problem in depth in a future work.

We recall that, although our simplified model cannot capture all the intricacies of the evolution of the Kuiper belt (e.g. important secular resonances produced by the presence of the other giant planets) an order of magnitude estimation is in place in order to motivate or discard the contribution of this mechanism in our planetary system.

As mentioned before, estimates of the mass of the CKB, vary between $0.008{\rm M}_{\oplus}<{M_{\it CKB}}<0.06{\rm M}_{\oplus}$ \citep{Trujillo01,Bernstein04,Fuentes08,Vitense10,Fraser14},
with most of them laying between $0.01{\rm M}_{\oplus}<{M_{\it CKB}}<0.03{\rm M}_{\oplus}$; since our generic disk does not limit itself to the cold component of the CKB nor accurately resemble the whole CKB (i.e. considering both the cold and hot components), we will proceed with the most conservative assumption in order to estimate the injection rate of comets in our solar system toy model, therefore we will make use of the results from our first disk mass simulation with $M_{DPs}=0.01$M$_\oplus$ (those who prefer larger masses can use the second or third masses presented in this work).

From panels \ref{sfig:b}, \ref{scros:b}, and \ref{sfreg:b} we can see that the total number of crossers by the end of the simulation is 19. We have studied each of these 19 crossers to see if any is locked in a resonance and thus could be resonant protected against a direct encounter with the giant planet; we found that none of them is well trapped inside the resonance (i.e. the resonant argument does not consistently librate, but it changes regime from circulating to librating), therefore, given enough time, all of them will have a close encounter with the Neptune-like giant. We can now estimate that the rate at which particles reach the Neptune-like neighborhood, as a direct result of the resupplying mechanism induced by DPs, is $\sim4\times10^{-12}$ yr$^{-1}$. In works from the literature, that consider the gravitational perturbations from the four giant planets of the solar system, the rate of escapes from the Kuiper belt region was found to be about 4 to 6$\times10^{-11}$ yr$^{-1}$ \citep{Duncan95,Levison97,Volk13}; our estimate is smaller by an order of magnitude. The same authors found that between 10-30\% of particles would become short-period comets after they first encounter Neptune, thus for our estimation we will use the middle value, 20\%. Although in this work we are only considering one giant planet and a cold disk, the inclusion of more giant planets would increase the number of resonances in the belt, increasing in turn the number of regions that can be repopulated by the secular effect, and enhancing the efficiency of the DPs (probably only by a small factor).

In our scenario, the total injection rate of new comets to the inner solar system depends on the total number of cometary nuclei in the size range of $\sim$ 1 to 10 km, specifically located in the CKB, between $\sim$ 38 and 50 au. This number is not well constrained by observations due to the intrinsic faintness of such objects. Nonetheless, many authors have tried to estimate the required number of objects in the source region to account for the observed population of ecliptic comets, assuming it remains in steady state. Most works have focused on estimating the number of cometary objects in the scattered disk larger than $\sim$2 km, since the scattered disk is the dominant source of ecliptic comets for those models, finding that reservoirs in the range $\sim4.4\times10^8$ to $\sim6\times10^9$ cometary nuclei can account for the observed steady-state ecliptic comet population, depending on the model \citep[e.g.][]{Levison97,Brasser13,Brasser15,Nesvorny17,Rickman17}. However, it is interesting to note that the most complete end-to-end simulations, those performed by Nesvorny et al., are still \emph{anemic} when accounting for the population of large comets by at least a factor of two, unless an evolution size-dependent model is assumed. 

Regarding the mass, several surveys suggest that the mass of the scattered disk could be similar to that of the CKB or up to a few times more massive \citep{Trujillo01,Bernstein04,Fuentes08,Vitense10}, with most of the values laying close to 1.5 times larger.

Regarding the population of cometary nuclei in the CKB: based on 4 Gyr simulations, \cite{Volk13} estimates that a source population of (1.7-3.5)$\times10^9$ objects larger than 1 km would be necessary to account for the number of Jupiter family comets (assuming steady state; if their only source was the CKB). Since the CKB accounts for approximately 40\% of the combined mass of the CKB plus the scattered disk, we will assume $(1.0 \pm 0.4) \times10^9$ cometary sized objects in the CKB, in order to estimate the injection rate produced purely by the DP-induced resupplying mechanism in the solar system. This number is still below the most recent constraints set by occultation surveys \citep{Schlichting09,Bianco10}.

By the previous arguments, an order of magnitude injection rate can be estimated as the product of the precedent factors (number of objects in the CKB, times the rate of escapes through the resonances, times the fraction of particles that become short period comets after first encountering Neptune), therefore, the mechanism described here can potentially add a rate of 7.6$\times10^{-4}$ comets/yr, to previous theoretical expectations. As a point of reference, a recent estimation of the rate required to account for the population of visible Jupiter family comets is (8.4$\pm$1.7)$\times10^{-3}$ comets/yr \citep{Rickman17}, therefore, the mechanism here proposed would account for $\sim 9\%$ of such rate; this fraction depends on several quantities that are poorly understood, and can easily go from 5\% up to 16\% (the latter occurs if we take the upper limit for the number of cometary nuclei in the reservoir and the lower limit in the estimate in the injection rate of short period comets), but this number could be even as high as 50$\%$ if the mass of the CKB is larger than the one we assumed.

Also, recent discoveries of large TNOs \citep[e.g.][]{Bannister16,Holman17} show that the number of DP-sized objects in the trans-Neptunian region is of the order of a few tens and, even when most of new discoveries involve dynamically hot objects, such large bodies are still able to affect the population of a cold disk, as we have previously shown by \citet{Munoz17}. We have looked for crossers in the 30 degree simulation from Mu\~noz-Guti\'errez et al., where we find that the efficiency drops by a factor of 2.1; however, the oblique encounters studied in that simulation are also similar to the interactions of the scattered disk with scattered DPs and thus, we can include the entire reservoir of cometary nuclei in the Kuiper belt that represents an increase of a factor of 2.5. Overall, this represents an increase of a factor of 1.18, when compared with our previous calculations. New members of the growing population of DP-sized objects in the solar system reaffirm the significance of the mechanism modeled here.

We do not claim this to be the rate at which new ecliptic comets are actually injected into the inner solar system. In fact, we would not be surprised if state of the art simulations of the solar system change this rate by a factor of two or more when compared to our generic model. Instead we show the potential importance of the mechanism proposed, besides the potential significance of the 2:1 MMR with Neptune as a source of new comets.

A complete and detailed numerical simulation, including all the four giant planets and the known trans-Neptunian DPs and largest objects, with an unbiased population of test particles that includes the diverse families of objects in the Kuiper belt, such as the cold CKB, the hot CKB, the scattered disk and the resonant populations, would be required to calculate accurately the injection rates of comets to the inner solar system. Such a simulation is beyond the scope of the present work.

\subsection{Injection Rates of Ecliptic Comets in Extrasolar Systems}
\label{Injection}

It is straightforward to extrapolate the existence of comets to extrasolar systems, specially considering that the exo-debris disks we know the best, such as the ones of Fomalhaut, Vega, HR4796, $\beta$Pictoris, $\epsilon$Eridani, and $\tau$Ceti, among others, are observed to be much more massive than our Kuiper belt. In fact, several studies have been presented already showing the highly likely presence of bodies from Ceres-size DPs to exo-comets in many systems \citep[e.g.][]{Weaver95,Movshovitz12,Bodman16,Rappaport18,Wyatt18}. In the near future, we will be able to study in a more statistical fashion processes that take comets to the inner planetary systems as we do now for the Kuiper belt. 

The injection of comets to the inner part of extrasolar systems has been explored as a mechanism to explain the observed exozodiacal clouds \citep{Bonsor12b}, which are thought to result from starlight re-radiated by dust remnants of exocomet tails (analogous to the way the inhomogeneities observed in the solar system Zodiacal Cloud are believed to originate from comet tails residuals). Also, from an astrobiological point of view, the number of comets in young planetary systems can be critical as they constitute a potential risk for the habitability of rocky exoplanets, inside the narrow region where life can develop.

Massive exoplanetary debris belts (some of which are observed to be orders of magnitude more massive than the one in the solar system) could contain larger numbers of cometary-sized objects; this, combined with their larger mass in DPs, even larger than those explored in this work, would elevate the significance of the resupplying mechanism for those systems.

In the case of our most massive disk, 0.25M$_\oplus$, the number of crossers at the end of the simulation is 592, which means that the rate at which particles reach the giant neighborhood is as much as $\sim 1\times10^{-10}$ yr$^{-1}$. From this and assuming similar conditions as for the the solar system (i.e. roughly 20\% of the escaped particles reaching the stage of ecliptic comets) and a number of objects in the debris disk 25 times larger than what we assumed for the CKB (i.e. $\sim 2.5\times10^{10}$), the injection rate turns out to be around 0.6 comets per year.

Naturally, the rate at which particles are passed from the outer system to the inner region will depend on the detailed configuration of giant planets in the system. For example, \cite{Bonsor12} have shown that the giant planets of the solar system are close to the optimal distribution in mass as well as spatially, to maximize the transfer rate of comets to the inner planetary system; this means that the $\sim20\%$ efficiency used earlier could be an overestimation for extrasolar systems.

Again, we do not claim this to be a number representative for any particular system, but we show the importance of the resupplying mechanism induced by DP-sized objects in driving the otherwise stable particles into perturbed
orbits, which could then become scattered by a giant planet. Also, we argue that the modeling of debris disks should consider the population of DP-sized objects if the secular dynamical effects are expected to be accounted for accurately.

\section{Conclusions}
\label{conc}

In this work we have explored a previously unstudied secular dynamical mechanism operating on debris disks: one induced by the presence of dozens of DP sized objects, that can potentially drive new cometary material into the strongest mean motion resonances of an interior giant planet. Once trapped in such resonances, the cometary nuclei can be destabilized either by slow chaotic diffusion or by a combination of this effect and perturbations from nearby massive DPs. The perturbed cometary nuclei have their eccentricities stirred enough to encounter the giant, potentially being sent into other dynamical families, such as a component analogue to the scattered disk or all the way down into the inner planetary system, where they would become ecliptic comets (given that their initial inclination is small).

The full dynamical path, from cometary nucleus in a debris disks to ecliptic comet, that we are introducing here, consists of several stages that can be summarized as follows
\begin{enumerate}
\item Replenishment: the repopulation of cometary nuclei into the resonances due to the secular interaction with DPs.
\item Stirring: the initial growth of the eccentricity (beyond $e_{lim}$) due to the resonant effect of the giant planet.
\item Crossers: the eccentricity continues growing until the periastron is comparable to the giant planet's orbit.
\item Direct interaction: after the first time that the cometary nucleus closely approaches the giant planet, it can severely change its path inwards or outwards.
\item Cascade: cometary nuclei going inwards can potentially interact with other interior giant planets.
\item Short period Comets: A few of the cometary nuclei will get stabilized by the most interior giant planet in cometary orbits of short period with almost constant Tisserand parameter.
\end{enumerate}

In this work, we have explored in detail the evolution of particles from stage 1 to stage 3, enough to demonstrate the proof of concept we are introducing in this work. Although stage 4 can be seen in the two most massive examples, in the lower mass examples it takes longer than 1 Gyr to set up and does not affect the proof of concept, furthermore, stages 5 and 6 strongly depend on the specific planetary system under consideration, and those lay outside the scope of this paper.

In order to motivate further studies, we present predictions based on a toy model of the solar system, where we find this effect to be present. If we consider work from the literature to quantify the later stages (4 to 6), we find that with enough DPs in the solar system (DPs in the range size between Mimas and Pluto, assuming many of the smaller ones are still to be discovered), this mechanism would be able to contribute with around 9\% of the injection rate of the new material required to sustain a steady-state population of ecliptic comets; this value depends on several poorly constrained quantities and can go from 5\% up to 16\%. Future dedicated surveys, like TAOS II that aims to characterize the population of the smaller bodies in the CKB \citep{Lehner16}, will be of great importance to test these figures in detail.

In extrasolar systems, the destabilizing mechanism explored here may account for the population of comets that contribute to the existence of exozodiacal clouds. Also, if a large number of new comets are being injected to the inner parts of extrasolar systems, they can pose a threat to the habitability of inner rocky worlds. Such rates still need to be calculated for particular systems in which a debris disk is accompanied by one or more giant planets; but in our most massive scenario, the rate of new comets is estimated to be extremely large: around 0.6 comets per year. In young massive active systems, this number could be in accordance with events analogous to the late heavy bombardment in the early solar system.

\acknowledgments

We acknowledge an anonymous referee for useful comments.
We acknowledge grant CONACyT Ciencia B\'asica 255167. AP acknowledges grant
DGAPA-PAPIIT IN109716. We acknowledge the use of the \emph{Atocatl} 
supercomputer at the Instituto de Astronom\'ia of the Universidad Nacional Aut\'onoma de M\'exico, where the simulations were performed.

\clearpage


\end{document}